\documentclass[review]{elsarticle}

\newcommand{\barb}{{\bar{b}}}
\newcommand{\barc}{{\bar{c}}}
\usepackage{verbatim}
\usepackage{graphicx}
\usepackage[usenames,dvipsnames,svgnames,table]{xcolor}
\usepackage[breaklinks=true,colorlinks=true,linkcolor=blue,urlcolor=blue,citecolor=blue]{hyperref}
\usepackage{rotating}
\usepackage{graphicx}
\usepackage{dcolumn}
\usepackage{bm}
\usepackage{epsfig}
\usepackage{hyperref}
\usepackage{ulem}
\usepackage{appendix}
\usepackage{mwe}
\usepackage{subfig}
\usepackage{lineno}
\expandafter\let\csname equation*\endcsname\relax
\expandafter\let\csname endequation*\endcsname\relax
\usepackage{amsmath}
\usepackage{geometry}
\newgeometry{vmargin={20mm}, hmargin={30mm,22mm}}

\begin{document}

\begin{frontmatter}
  
\title{Bottomonia production in p+p collisions under NRQCD formalism}

\author[NPD]{Vineet Kumar\corref{mycorrespondingauthor}}
\cortext[mycorrespondingauthor]{Corresponding author}
\ead{vineetk@barc.gov.in}
\author[UOC]{Kinkar Saha}
\author[NPD,HBNI]{Prashant Shukla}
\author[UOC]{Abhijit Bhattacharyya}

\address[NPD]{Nuclear Physics Division, Bhabha Atomic Research Centre, Mumbai 400085, India}
\address[UOC]{Department of Physics, University of Calcutta, 92, A. P. C. Road Kolkata-700009, India}
\address[HBNI]{Homi Bhabha National Institute, Anushakti Nagar, Mumbai 400094, India}

\date{\today}

\begin{abstract}
  In this work, we present the calculation of the production cross sections of bottomonia
  states using Non-Relativistic Quantum Chromodynamics (NRQCD) formalism.
  The direct production cross-section of a resonance can be factorised in terms of
  short distance Quantum Chromodynamics (QCD) cross sections and long distance matrix
  elements (LDMEs) under NRQCD.  We use a large set of measured $\Upsilon$(nS) production
  data at Tevatron and LHC energies in both central and forward rapidity regions to extract 
  the LDMEs with better precision.
  The feed down contributions from the higher states including the $\chi_{b}$(3P) state are taken into
  account for the LDME extraction.
  The formalism provides a good description of the bottomonia data in wide transverse momentum
  range at different collision energies. 
\end{abstract}


%
%
%
%
%
\end{frontmatter}

\section{Introduction}

  Quantum Chromodynamics (QCD) describes the strong interaction among the
quarks and gluons via perturbative calculations utilising its property called asymptotic freedom.
On the other hand, these quarks and gluons are confined inside hadrons which are the colour singlet states.
Confinement is a purely non-perturbative phenomenon which is not very well understood yet. 
The study of quarkonia ($Q\bar{Q}$) serves as an effective 
tool to look at  both of these perturbative and non-perturbative aspects of QCD.
The quarkonia states differ from most other hadrons due to the small velocity, $v$ of the massive
constituents and thus can be treated using non-relativistic formalism~\cite{Povh:1995mua,Ikhdair:2005jf}. 
In a simple picture, one can think of a quarkonium as a heavy quark pair ($Q\bar{Q}$) bound
in a colour singlet state by some effective potential interaction, where the constituents are 
separated by distances much smaller than $1/\Lambda_{\rm QCD}$ where $\Lambda_{\rm QCD}$
is the QCD scale. This interaction gets screened 
in the presence of a deconfined medium like Quark Gluon Plasma (QGP), causing 
the bound state to melt away and thus the quarkonia yields are suppressed in the 
heavy ion collisions. This makes quarkonia an important probe of QGP. However cold nuclear matter 
effects such as modification of parton distribution functions of nucleons inside nucleus also 
affect their yields.
 There have been immense experimental~\cite{Sirunyan:2017isk,Sirunyan:2018nsz,Acharya:2019iur,Acharya:2018mni}
and theoretical works~\cite{Strickland:2011mw,Song:2011nu,Kumar:2014kfa,Kumar:2019xdj} on
quarkonia modifications in PbPb collisions for which understanding of quarkonia
production in pp collisions is an important prerequisite.

 The massive quarks (with $m_c\sim 1.6$ GeV/$c^2$, $m_b\sim 4.5$ GeV/$c^2$) are produced
in initial stages in hadronic collision with high momentum transfer and thus
can be treated perturbatively~\cite{Nason:1989zy}. The emergence of quarkonia
out of the two massive quarks, on the other hand can only be described non-perturbatively using different
models~\cite{Bodwin:1994jh,Brambilla:2014jmp}.
The Colour Singlet Model (CSM)~\cite{Einhorn:1975ua,Berger:1980ni},
Colour Evaporation Model (CEM)~\cite{Fritzsch:1977ay,Amundson:1995em}, the Fragmentation Scheme and 
the NRQCD factorisation formalism are some of the well established models for quarkonia production.
 In the framework of CSM, the $Q\bar{Q}$ pair, eventually evolving into the quarkonium,
is assumed to be in Colour Singlet (CS) state and that has spin and 
angular momentum same as that of quarkonium.
Apart from comprising of the CSM, the NRQCD factorisation approach incorporates 
the Colour Octet (CO) states as well.

In the formalism of the NRQCD factorisation approach, the evolution probability of $Q\bar{Q}$
pair into a state of quarkonium is expressed as matrix elements of NRQCD operators expanded
in terms of heavy quark velocity $v$ (for $v\ll$1)~\cite{Bodwin:1994jh}.
The factorisation formulae were then used to calculate production cross-sections
and decay rates of quarkonia states.
The full structure of the $Q\bar{Q}$ Fock space
is considered and spanned by $n$=$^{2s+1}L_J^{[a]}$ state where $s$
is the spin, $L$ is the orbital angular momentum, $J$ is the total angular momentum
and $a$ (colour multiplicity) = 1 for CS and 8 for CO states. 
The produced CO states of $Q\bar{Q}$ pair at short distances emerge as 
CS quarkonia by emitting soft gluons non-perturbatively.
The short distance cross-sections are obtained theoretically
using methods of perturbative QCD (pQCD). The long distance matrix elements
(LDME) that correspond to the probability of 
$Q\bar{Q}$ pair to emerge as quarkonium are extracted by fitting the measured cross-section
data.

 There have been several works on bottomonia production based on
NRQCD formalism. In Ref.~\cite{Domenech:1999qg}, a Monte Carlo framework has first
been employed with CO mechanism for inclusive bottomonia production and few
NRQCD CO matrix elements for $\Upsilon$(1S) have been extracted at the Tevatron energy. 
The study has been extended to the whole $\Upsilon$(nS) family in Ref.~\cite{Domenech:2000ri}
to find CO matrix elements using CDF measurements at Tevatron.
In Ref.~\cite{Brateen:PRD2001} the CO matrix elements are obtained for $\Upsilon$(nS) family
and the feed downs from $\chi_{b}$(1P) and $\chi_{b}$(2P) to $\Upsilon$(1S) have been 
considered.
 In Ref.~\cite{Gong:2010bk}, the $\Upsilon$ production has been obtained via
S-wave CO states calculated at Next to Leading Order (NLO). The LDMEs are obtained
by fitting the Tevatron data. The ratios of NLO to LO total cross-sections
have been obtained at Tevatron and LHC energies. Polarisation of inclusive
$\Upsilon$ has been obtained albeit with large uncertainties.
In Ref.~\cite{Sharma:2012dy} both CS and CO states along with
feed down contributions from higher states have been considered to study the
quarkonia yields for RHIC and LHC energies.
Using Collins-Soper-Sterman (CSS) formalism, an extension of the NRQCD prediction
has been carried forward for heavy quarkonium production
at low $p_T$ by considering soft gluon resummation at all orders in Ref.~\cite{Sun:2012vc}.

Both production and polarisation of $\Upsilon$(nS) at NLO have been discussed in 
Ref.~\cite{Gong:2013qka} within the framework of NRQCD. The CO matrix elements are obtained
by fitting with experimental data. The study is updated in Ref.~\cite{Feng:2015wka} by considering
feed down from $\chi_{bJ}$(mP) states in $\Upsilon$(nS) production. The yields and
polarisations of $\Upsilon$(nS) measured at Tevatron and LHC are well explained by this work.
The NLO study in Ref.~\cite{Han:2014kxa} describes the yields and polarisations of
$\Upsilon$(nS) at LHC which includes feed down contributions from
higher states. {\color{black} Ref.~\cite{Feng:2020cvm} gives complete analysis of the polarization parameters
of $\Upsilon$(nS) at QCD next-to-leading order in both the helicity and Collins-Soper frames.}
In Ref.~\cite{Yu:2017pot}, production cross-section for $\Upsilon$(nS),
$\chi_{bJ}$, $\eta_b$ and $h_b$ have been calculated using NRQCD, as produced in hard
photo production and fragmentation processes at LHC energies. 

A LO NRQCD analyis is useful as it is straightforward and unique and once the parameters are
obtained by fitting over large datasets it has excellent predictability power for unknown cross
sections. The short distance QCD cross-sections calculation techniques at NLO are not unique.
Moreover the different components of pQCD NLO cross sections are not available in
public domain. Many NLO analysis do not include the feed down contribution from the higher
states. It is shown that there is a large difference amoung the LDMEs obtained by different
analysis at NLO. In this paper, the LO NRQCD calculations for the differential production
cross-sections of $\Upsilon$ states in p+p collisions have been presented. Our work includes
most up to date datasets and feeddown contributions.
{\color{black} We have given an
estimate of uncertainties in the LDMEs due to enhancement of CS quarkonia cross-section by a
factor of two expected from the NLO corrections, only slight changes appear in the CO quarkonia
cross-section when the NLO QCD corrections are included~\cite{Gong:2010bk,Gong:2008hk}.}

The NRQCD formalism is described
briefly in Section~\ref{sec:formalism}. 
A large set of data from Tevatron~\cite{Acosta:2001gv} and
LHC~\cite{LHCb:2012aa,Khachatryan:2015qpa,Aad:2012dlq,Chatrchyan:2013yna,Sirunyan:2017qdw} 
is used to extract the LDMEs required for the $\Upsilon$ production and then results are
presented in Section~\ref{sec:results}. A comparison of the obtained LDMEs with the
previous NRQCD studies both at LO and NLO has been made.    
The summary of our findings are discussed in Section~\ref{sec:summary}.


\begin{table*}
  \centering
  \caption{Necessary and pertinent branching fractions for bottomonia family~\cite{Han:2014kxa,Zyla:2020zbs}.}
  \footnotesize
  \begin{tabular*}{\textwidth}{@{\extracolsep{\fill}}lrrrrrrrrl@{}}
    \hline
    \hline
    & & & & & & & & &\\
    Meson from & & & & & Meson to & & & & \\ \\
    \hline 
    &$\Upsilon$(3S) &$\chi_{b0}$(2P) &$\chi_{b1}$(2P) &$\chi_{b2}$(2P) &$\Upsilon$(2S) &$\chi_{b0}$(1P) &$\chi_{b1}$(1P) &$\chi_{b2}$(1P) &$\Upsilon$(1S)\\
    \hline
    \hline \\
    $\chi_{b0}$(3P) &0.005 & & & &0.002 & & & & 0.002 \\ \\
    $\chi_{b1}$(3P) &0.104 & & & &0.037 & & & & 0.038 \\ \\
    $\chi_{b2}$(3P) &0.061 & & & &0.019 & & & & 0.019 \\ \\
    $\Upsilon$(3S) & & 0.131 &0.126 & 0.059 & 0.199 & 0.003 & 0.0017 & 0.019 & 0.066 \\ \\
    $\chi_{b0}$(2P) & & & & &0.014 & & & & 0.004 \\ \\
    $\chi_{b1}$(2P) & & & & &0.199 & & & & 0.092 \\ \\
    $\chi_{b2}$(2P) & & & & &0.106 & & & & 0.070 \\ \\
    $\Upsilon$(2S) & & & & & & 0.038 & 0.0715 & 0.069 & 0.260 \\ \\
    $\chi_{b0}$(1P) & & & & & & & & & 0.019 \\ \\
    $\chi_{b1}$(1P) & & & & & & & & & 0.352 \\ \\
    $\chi_{b2}$(1P) & & & & & & & & & 0.180 \\ \\
    \hline
    \hline
  \end{tabular*}
  \label{BRUpsilon}
\end{table*}
\normalsize
\begin{table*}
  \centering
  \caption{CS and CO elements for $\Upsilon$ family, obtained theoretically/extracted using experimental data~\cite{Sharma:2012dy,Feng:2015wka}.}
  \footnotesize
  \begin{tabular*}{\textwidth}{@{\extracolsep{\fill}}lrrrrl@{}}
    \hline
    \hline
    & & \\ 
    Direct Contributions & Feed down contributions & Feed down contributions \\
    & from higher s-wave states & from higher p-wave states \\ \\
    \hline 
    & & \\
    $M_L(b\bar{b}([^3S_1]_1)\rightarrow\Upsilon(3S))$ & $M_L(b\bar{b}([^3S_1]_1)\rightarrow\Upsilon(3S,2S))$
    & $M_L(b\bar{b}([^3P_0]_1)\rightarrow\chi_{b0}(1P))$ \\
    =4.3 ${\rm GeV^3}$ & =4.3, 4.5 ${\rm GeV^3}$ & =0.100$m_b^2$ ${\rm GeV^3}$ \\ \\
    
    $M_L(b\bar{b}([^3S_1]_1)\rightarrow\Upsilon(2S))$ & $M_L(b\bar{b}([^3S_1]_8)\rightarrow\Upsilon(3S,2S))$
    & $M_L(b\bar{b}([^3S_1]_8)\rightarrow\chi_{b0}(1P))$ \\
    =4.5 ${\rm GeV^3}$ & & =0.0094 ${\rm GeV^3}$ \\ \\
    
    $M_L(b\bar{b}([^3S_1]_1)\rightarrow\Upsilon(1S))$ & $M_L(b\bar{b}([^1S_0]_8)\rightarrow\Upsilon(3S,2S))$
    & $M_L(b\bar{b}([^3P_0]_1)\rightarrow\chi_{b0}(2P))$ \\
    =10.9 ${\rm GeV^3}$ &  & =0.100$m_b^2$ ${\rm GeV^3}$ \\ \\
    
    $M_L(b\bar{b}([^3S_1]_8)\rightarrow\Upsilon(nS))$ & $M_L(b\bar{b}([^3P_0]_8)\rightarrow\Upsilon(3S,2S))$
    & $M_L(b\bar{b}([^3S_1]_8)\rightarrow\chi_{b0}(2P))$ \\
    &  & =0.0109 ${\rm GeV^3}$ \\ \\
    
    $M_L(b\bar{b}([^1S_0]_8)\rightarrow\Upsilon(nS))$ & $M_L(b\bar{b}([^3P_1]_8)\rightarrow\Upsilon(3S,2S))$ &$M_L(b\bar{b}([^3P_0]_1)\rightarrow\chi_{b0}(3P))$ \\
    & =3$M_L(b\bar{b}([^3P_0]_8)\rightarrow\Upsilon(3S,2S))$ &=0.100$m_b^2$ ${\rm GeV^3}$  \\ \\
    
    $M_L(b\bar{b}([^3P_0]_8)\rightarrow\Upsilon(nS))$ & $M_L(b\bar{b}([^3P_2]_8)\rightarrow\Upsilon(3S,2S))$ &$M_L(b\bar{b}([^3S_1]_8)\rightarrow\chi_{b0}(3P))$ \\
    & =5$M_L(b\bar{b}([^3P_0]_8)\rightarrow\Upsilon(3S,2S))$ &=0.0069 ${\rm GeV^3}$ \\ \\
    
    $M_L(b\bar{b}([^3P_1]_8)\rightarrow\Upsilon(nS))$ & & \\
    3$M_L(b\bar{b}([^3P_0]_8)\rightarrow\Upsilon(nS))$ & &  \\ \\
    
    $M_L(b\bar{b}([^3P_2]_8)\rightarrow\Upsilon(nS))$ & & \\
    5$M_L(b\bar{b}([^3P_0]_8)\rightarrow\Upsilon(nS))$ & &  \\ \\
    \hline
    \hline
  \end{tabular*}
  \label{CSCO}
\end{table*}
\normalsize

\section{Bottomonia production in p$+$p collisions}
\label{sec:formalism}
In order to study heavy quarkonium yield, the NRQCD framework serves as an
efficient theoretical tool. The processes that govern the differential
production of heavy mesons like bottomonium, as functions of $p_T$ are mostly
2$\rightarrow$2 operations. These processes can be denoted generically by 
$a+b\rightarrow \Upsilon +X$, where $a$ and $b$ are the incident light partons,
$\Upsilon$ is the heavy meson and $X$ is final state light parton.
The double differential cross-section as a function of $p_T$ and rapidity ($y$) of 
the heavy meson can be written as~\cite{Kumar:2016ojy},
\begin{eqnarray}
\nonumber
 \frac{d^2\sigma^{\Upsilon} }{dp_T dy} &=& \sum_{a,b} \int_{x_a^{min}}^1 dx_a G_{a/p}(x_a,\mu_F^2)
 G_{b/p}(x_b,\mu_F^2) \\
 &\times& 2p_T\frac{x_a x_b}{x_a - \frac{m_T}{\sqrt{s}}e^y}\frac{d\sigma}{d\hat{t}}
 \label{eq4}
\end{eqnarray}
where, $G_{a/p}$($G_{b/p}$) are the colliding parton $(a(b))$ distribution functions in
the incident protons. They depend on the fractions $x_a$($x_b$), of the total momentum
carried by the incident partons and the scale of factorisation $\mu_F$.
Here $\sqrt{s}$ represents the total center of mass energy of the pp system and $m_T~(=\mu_F)$ stands for
the transverse mass, $m_T^2=p_T^2 + M^2$ of the quarkonium. {\color{black}The one loop $\alpha_{\rm s}$(Q$^{2}$) is used
in the calculations and value of the Q$^{2}$ is taken equal to the square of the scale of factorisation ($\mu_F$).}
The relation between $x_a$ and $x_b$ and the expression for $x_a^{min}$ are given in our earlier
work~\cite{Kumar:2016ojy}.
 The ${d\sigma}/{d\hat{t}}$ in Eq.~\ref{eq4} is the parton level cross-section and is
defined as~\cite{Bodwin:1994jh},
\begin{equation}
  \frac{d\sigma}{d\hat{t}} = \frac{d\sigma}{d\hat{t}}(ab\rightarrow Q\bar{Q}(^{2s+1}L_J)+X)
  M_L(Q\bar{Q}(^{2s+1}L_J)\rightarrow \Upsilon)
  \label{eq6}
\end{equation}
The first term in RHS is the short distance contribution, that corresponds to the $Q\bar{Q}$
pair production in specific colour and spin configuration and is calculable using 
perturbative QCD (pQCD)~\cite{Brateen:PRD2001,Baier:1983va,Humpert:1986cy,Gastmans:1987be,Cho:1995vh,Cho:1995ce}.
The other term in the RHS of Eq.(\ref{eq6}) is the Long Distance Matrix Element (LDME)
and refers to the probability of the $Q\bar{Q}$ state to convert into a quarkonium state.
They are determined by contrasting with experimental observations.

The NRQCD formalism provides an adequate procedure to estimate a quantity as an expansion in 
heavy quark relative velocity, $v$ inside $Q\bar{Q}$ bound state. The LDME in Eq.(\ref{eq6})
do scale with definitive power in $v$. The quarkonium yield depends on the $^3S_1^{[1]}$ 
and $^3P_J^{[1]}$(J=0,1,2) CS states and $^1S_0^{[8]}$, $^3S_1^{[8]}$ and $^3P_J^{[8]}$
CO states in the limit $v\ll 1$.
The superscripts in square brackets represent the colour structure of the bound state,
1 for the CS and 8 for the CO. The direct production cross-section for 
$\Upsilon$ in differential form can thus be expressed as the sum of all contributions,
\begin{eqnarray}
d\sigma(\Upsilon(nS)) &=& d\sigma(Q\overline{Q}([^3S_1]_{1}))
                   M_{L}(Q\bar{Q}([^3S_1]_{1})\rightarrow \Upsilon(nS)) \nonumber \\
                &+& d\sigma(Q\overline{Q}([^1S_0]_{8}))
                   M_{L}(Q\bar{Q}([^1S_0]_{8})\rightarrow \Upsilon(nS)) \nonumber \\ 
                &+& d\sigma(Q\overline{Q}([^3S_1]_{8}))
                   M_{L}(Q\bar{Q}([^3S_1]_{8})\rightarrow \Upsilon(nS)) \nonumber \\
                &+& d\sigma(Q\overline{Q}([^3P_0]_{8}))
                   M_{L}(Q\bar{Q}([^3P_0]_{8})\rightarrow \Upsilon(nS))  \nonumber \\
                &+& d\sigma(Q\overline{Q}([^3P_1]_{8}))
                   M_{L}(Q\bar{Q}([^3P_1]_{8}) \rightarrow \Upsilon(nS)) \nonumber \\
                &+& d\sigma(Q\overline{Q}([^3P_2]_{8}))
                   M_{L}(Q\bar{Q}([^3P_2]_{8})\rightarrow \Upsilon(nS)) \nonumber \\
                &+& ...
\label{eq8}
\end{eqnarray}
The dots include terms having contributions in higher powers of $v$.

\begin{figure}[!h]
\centering
  \includegraphics[width=0.49\textwidth]{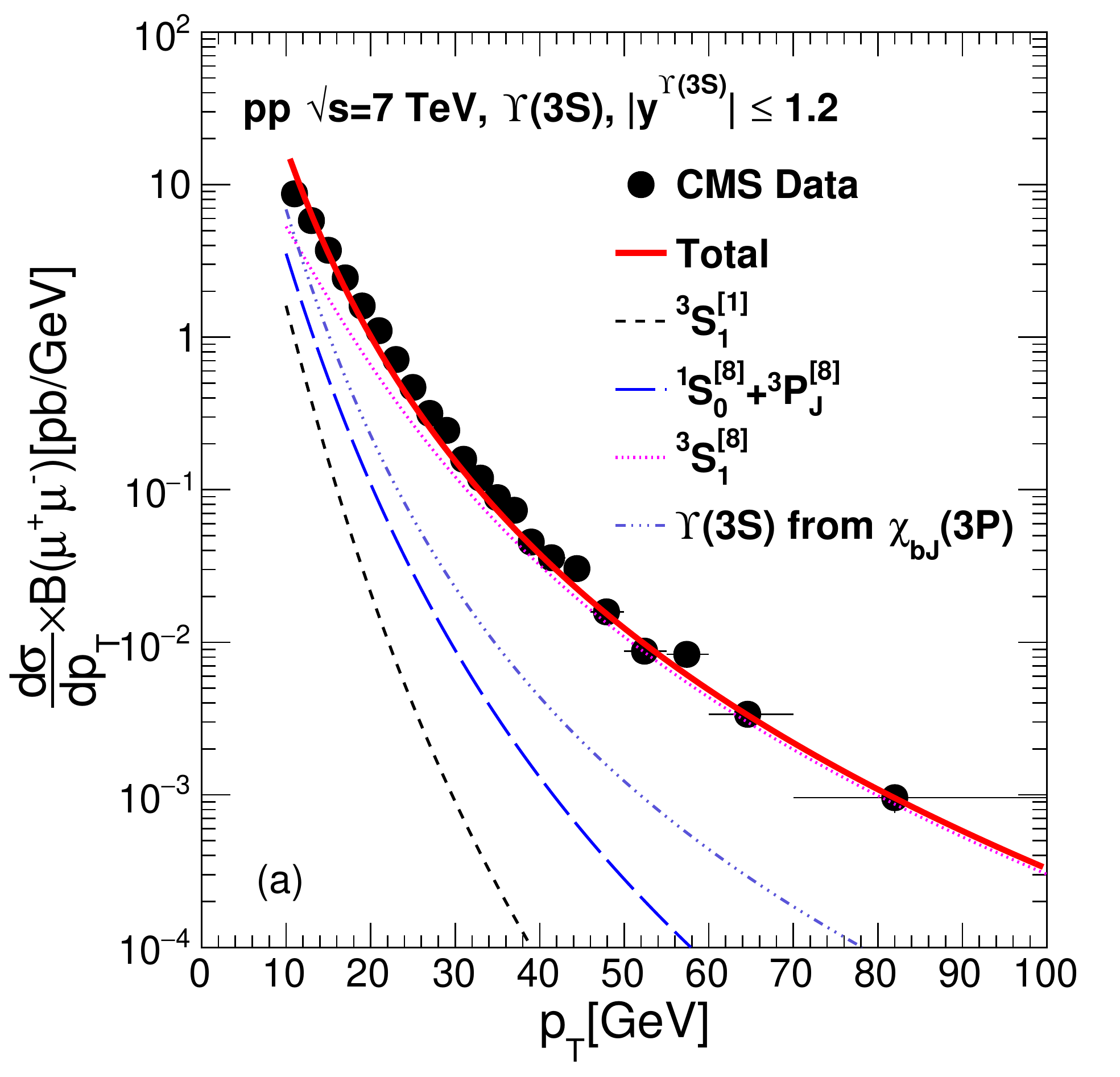}
  \includegraphics[width=0.49\textwidth]{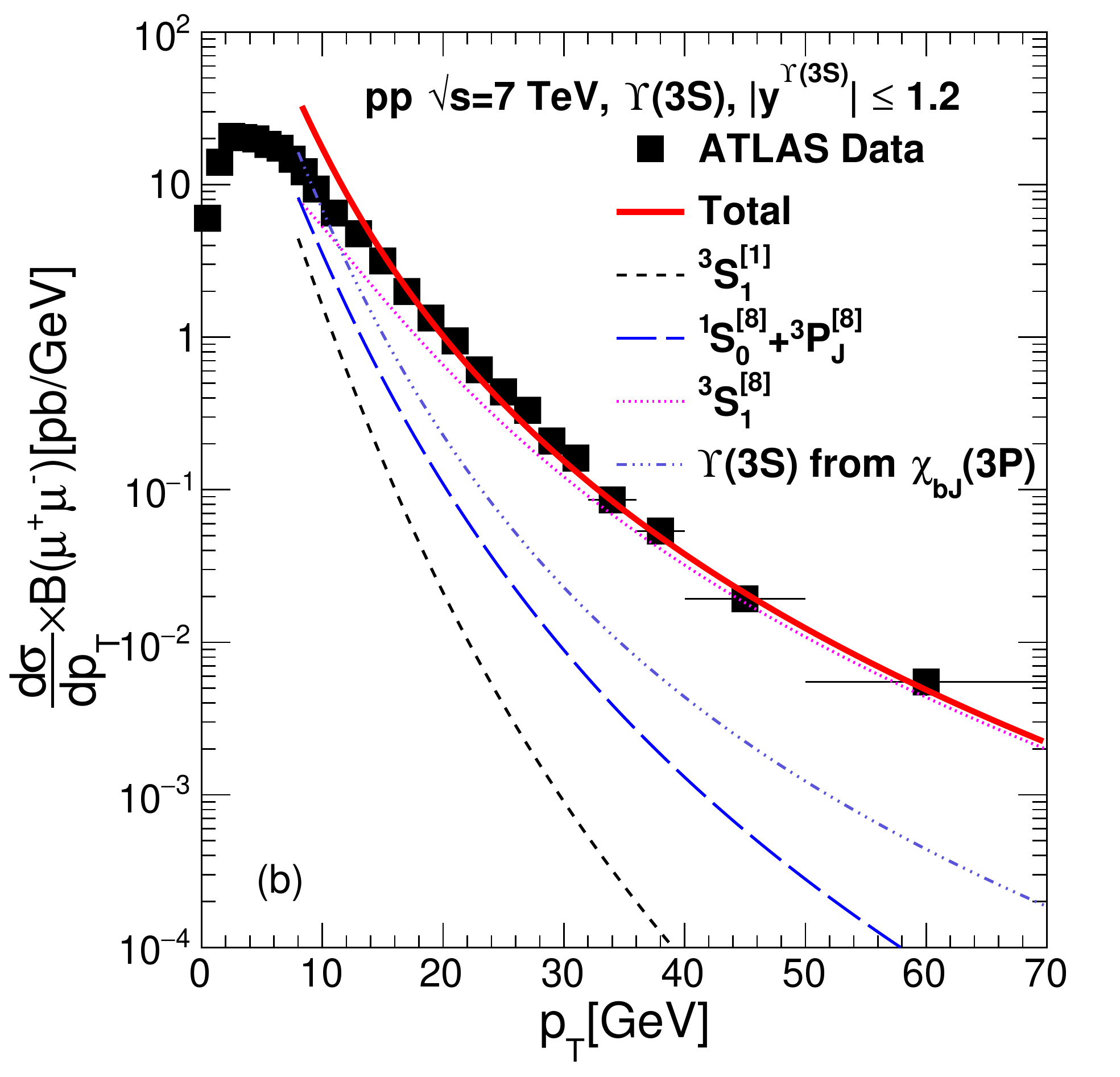} 
 \caption{\small{The NRQCD calculations of production cross-section of $\Upsilon$(3S) in p+p collisions at 
   $\sqrt{s}$ = 7 TeV in central rapidities, as a function of transverse momentum compared with the measured data 
   at CMS~\cite{Khachatryan:2015qpa} and ATLAS~\cite{Aad:2012dlq} experiment.} }
  \label{Fig:SigmaY3SCMS}
\end{figure}

The contributions from CS-$[^3P_J]_1$ and CO-$[^3S_1]_8$ states are in the same order
of $v$ for the p-wave bound states, $\chi_{b}(nP)$. The angular momentum barriers of the p-wave
states are held responsible for that to happen and thereby making them important enough
to be considered. The differential cross-section for $\chi_b$ production
henceforth is given by,
\begin{eqnarray}
 d\sigma(\chi_{bJ}(1P)) &=& d\sigma(Q\overline{Q}([^3P_J]_{1}))
                   M_{L}(Q\bar{Q}([^3P_J]_{1})\rightarrow \chi_{bJ}(1P)) \nonumber \\
                &+& d\sigma(Q\overline{Q}[^3S_1]_{8}))
                   M_{L}(Q\bar{Q}([^3S_1]_{8})\rightarrow \chi_{bJ}(1P))  \nonumber \\
                &+& ...
\label{eq9}
\end{eqnarray}

The experimental observations of $\Upsilon$ production at LHC energies, not only have contributions from
direct yield, but also consist of feed downs from decay of heavier bottomonia states.
The corresponding branching fractions are
provided in Table~\ref{BRUpsilon}.

\begin{figure}[!h]
\centering
  \includegraphics[width=0.49\textwidth]{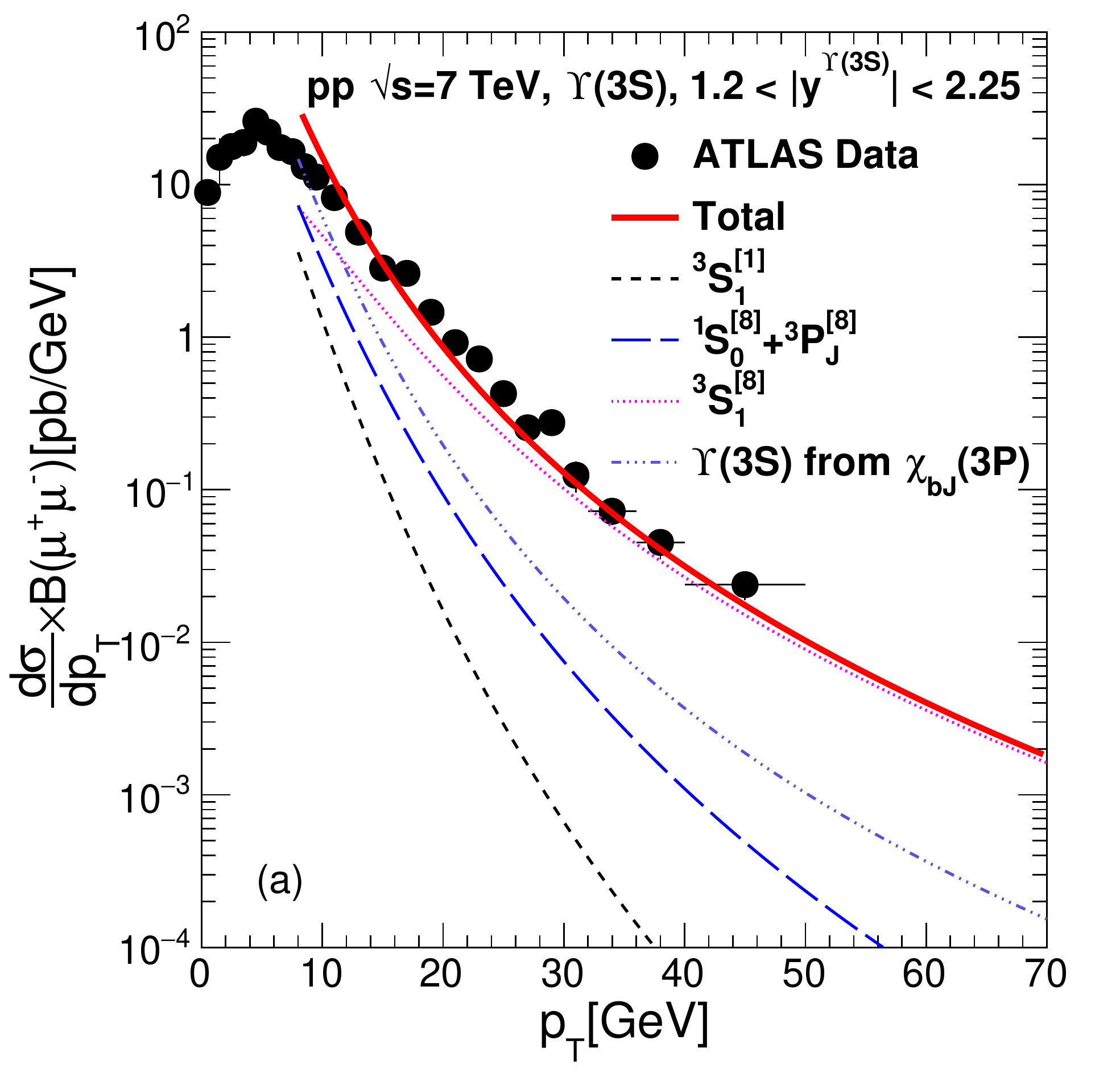}
  \includegraphics[width=0.49\textwidth]{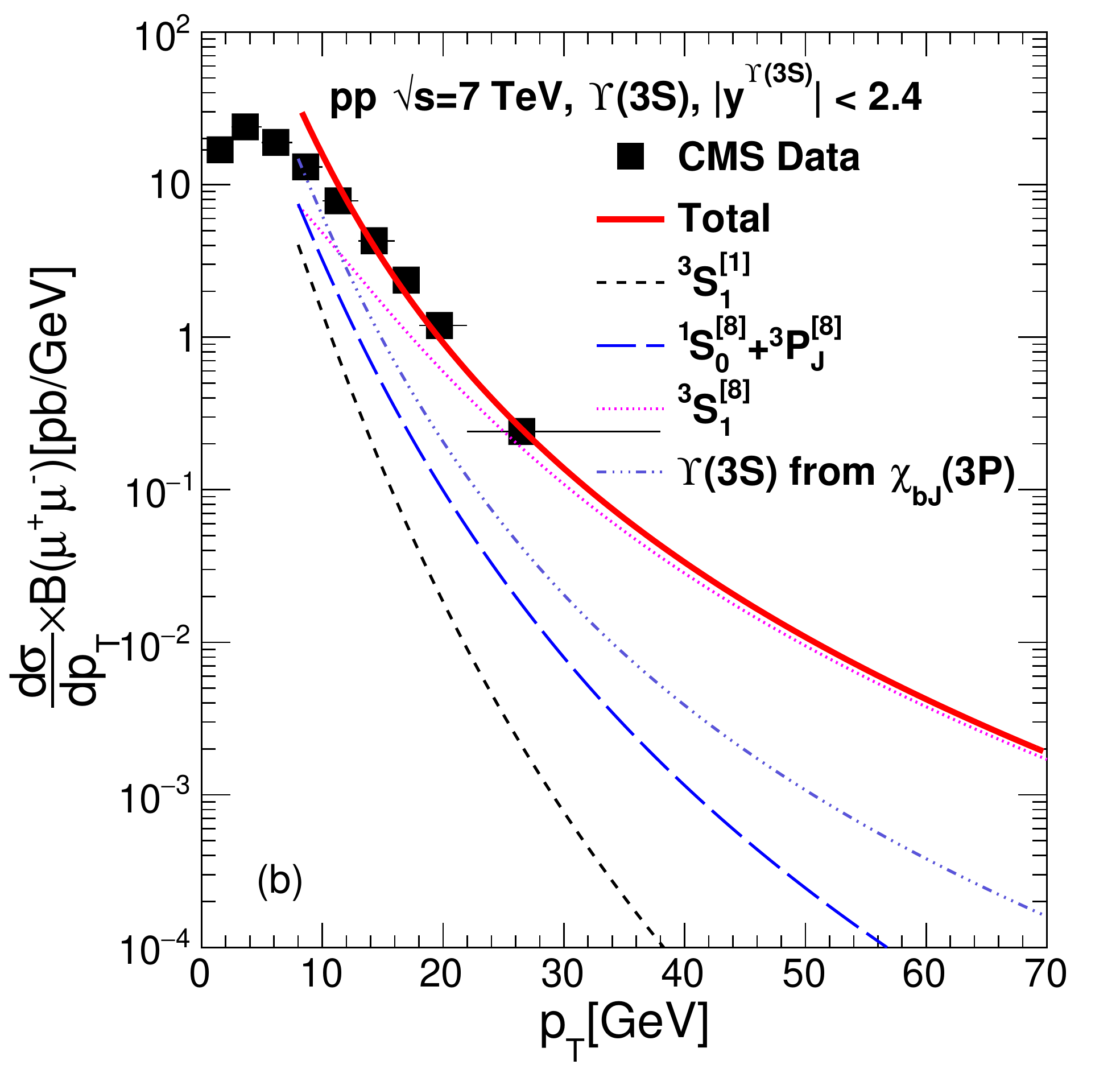} 
 \caption{\small{The NRQCD calculations of production cross-section of $\Upsilon$(3S) in p+p collisions at 
   $\sqrt{s}$ = 7 TeV in forward rapidities, as a function of transverse momentum compared with the measured data 
   at ATLAS~\cite{Aad:2012dlq} and CMS~\cite{Chatrchyan:2013yna} experiments. }}
  \label{Fig:SigmaY3SCMS_forwardRap}
\end{figure}

We require both CS and CO matrix elements in order to get theoretical
predictions for the production of bottomonia at the Tevatron and LHC energies.
The corresponding expressions and
numerical values for CS states are obtained from Ref.~\cite{Brateen:PRD2001}.
The CO states, on the other hand, cannot be directly connected to the non-relativistic
wavefunctions of heavy mesons,
as these are associated with a higher Fock state. Experimentally measured data sets are 
therefore employed to obtain them as in Refs.~\cite{Brateen:PRD2001,Cho:1995vh,Cho:1995ce}. 
The CS operators along with their theoretical values
and the CO operators to be fitted are listed in Table~\ref{CSCO},
where, $n$=1,2,3. For the CO elements related to p-wave states, needed as the 
feed down contributions, we have used values obtained by Ref.~\cite{Sharma:2012dy,Feng:2015wka} for the 
present purpose. In our calculations, we have used
CT14LO parametrisation~\cite{Hou:2019efy} for parton distribution functions and 
the bottom quark mass $m_b$ is taken to be 4.88 GeV.
 The short distance cross-sections for $[^1S_0]_8$ and $[^3P_J]_8$ states having similar 
$p_T$ dependencies, the corresponding distributions become sensitive upto a linear combination
 of their LDMEs. We therefore take resort to a linear combination following
 Ref.~\cite{Kumar:2016ojy} as,

\begin{equation*}
\begin{split}
  & M_L(b\bar{b}([^1S_0]_8,[^3P_0]_8)\rightarrow\Upsilon(nS)) = \\
  &\frac{M_L(b\bar{b}([^1S_0]_8)\rightarrow\Upsilon(nS))}{5} +\frac{3 M_L(b\bar{b}([^3P_0]_8)\rightarrow\Upsilon(nS))}{m_b^2}.\\
\end{split}
\end{equation*}
\begin{figure}
  \centering
  \includegraphics[width=0.49\textwidth]{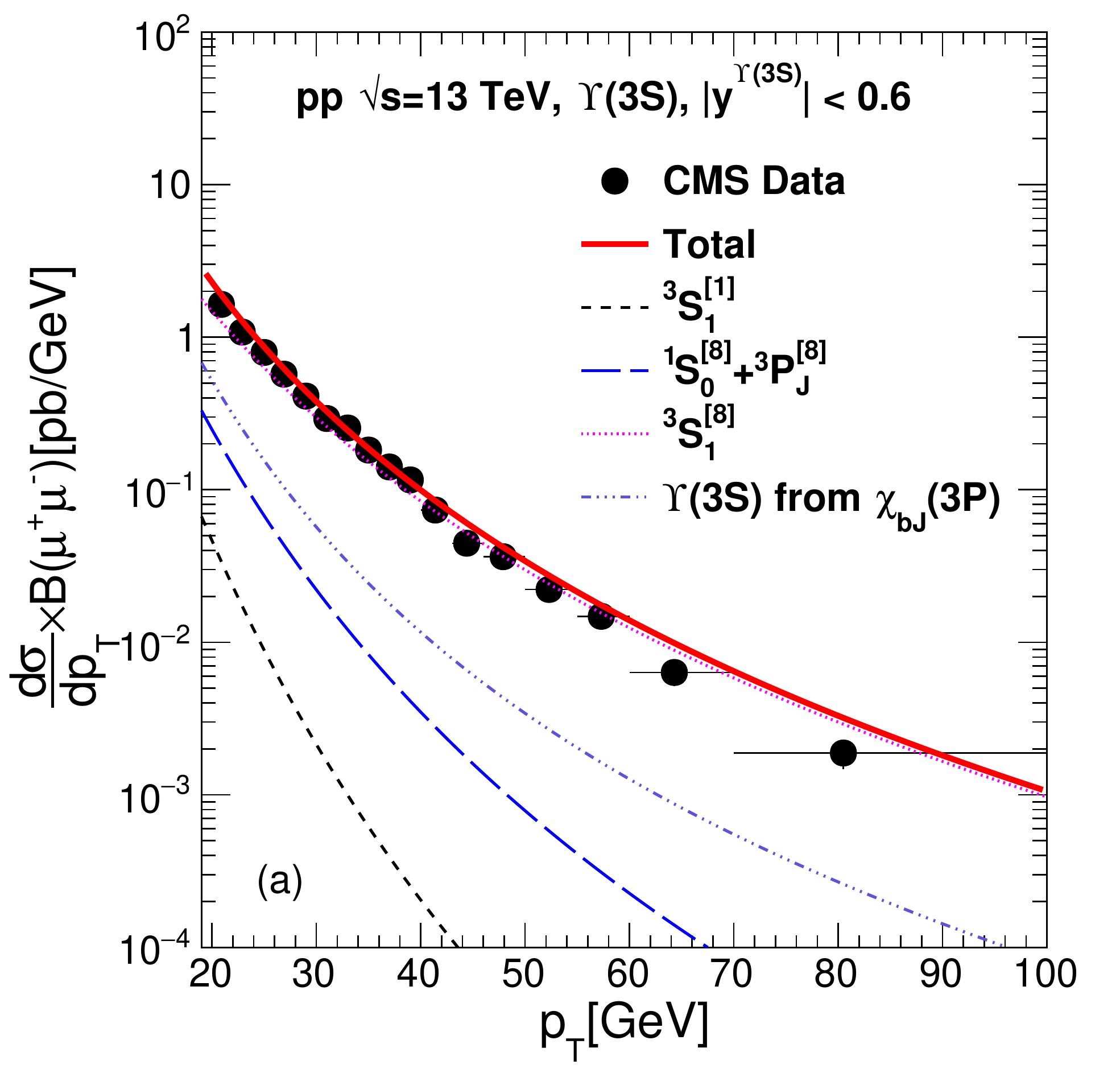}
  \includegraphics[width=0.49\textwidth]{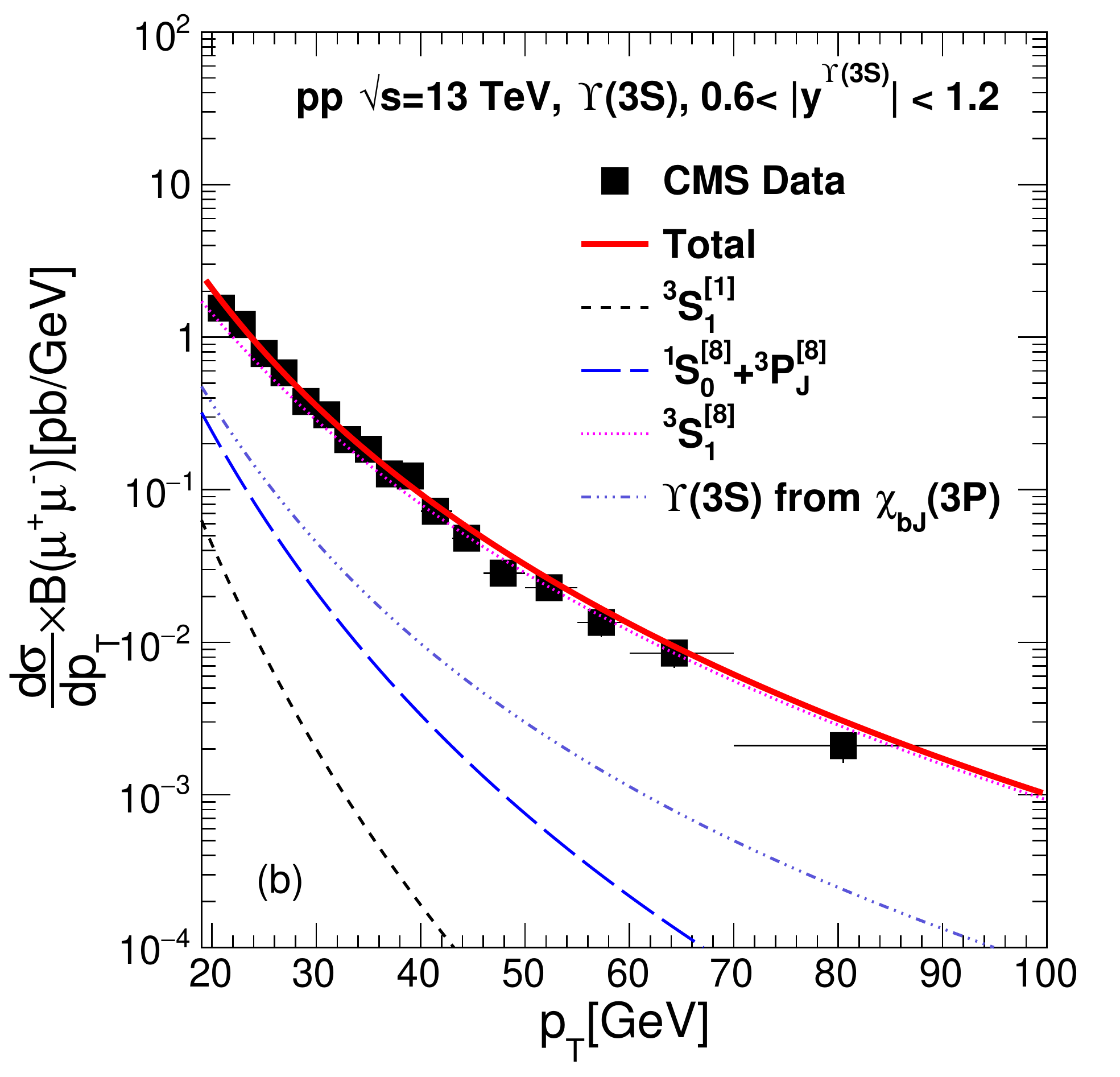}
  \includegraphics[width=0.49\textwidth]{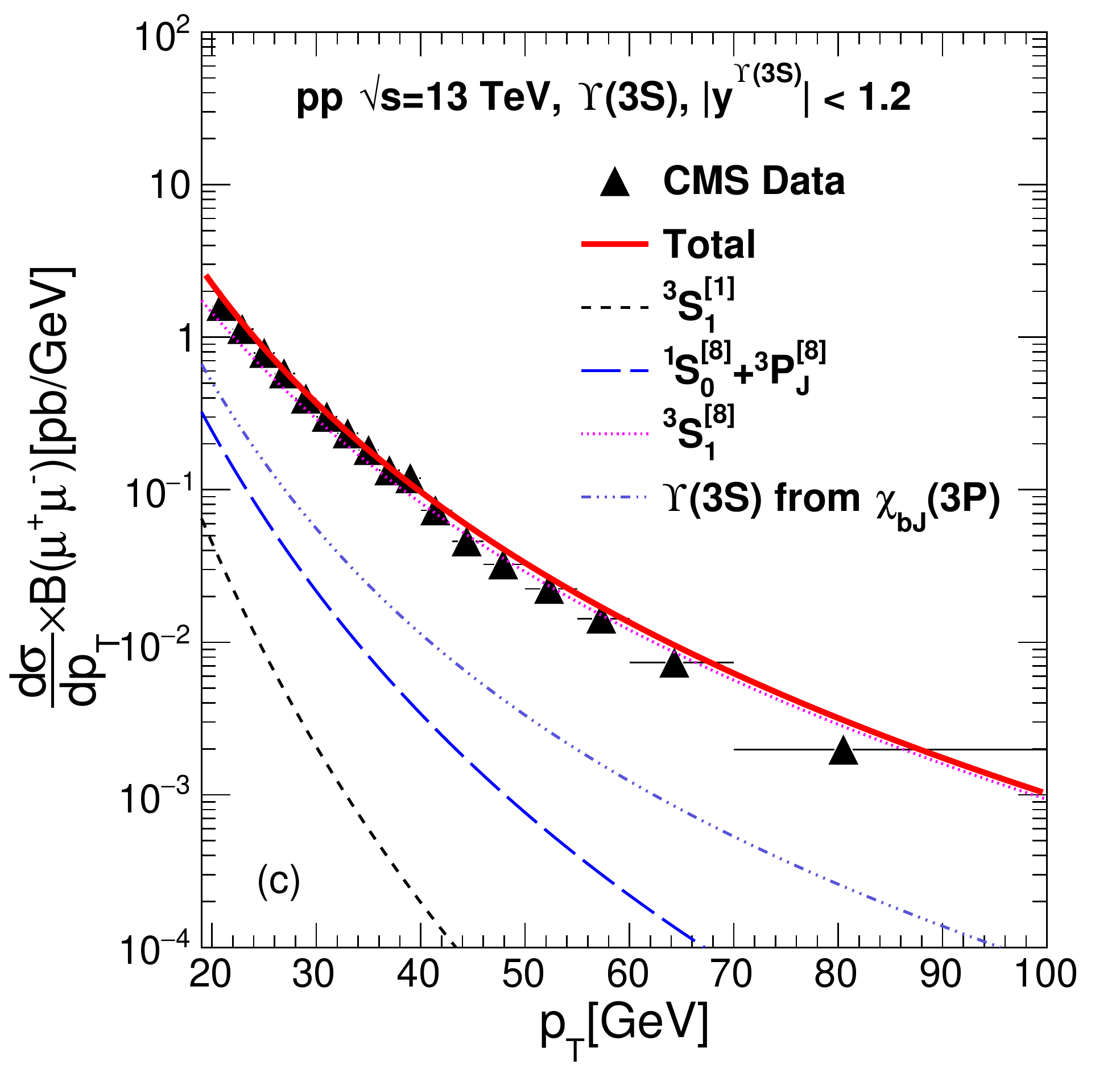}
  \caption{\small{The NRQCD calculations of production cross-section of $\Upsilon$(3S) in p+p collisions at 
    $\sqrt{s}$ = 13 TeV in central and forward rapidities, as a function of transverse momentum compared with the measured data 
    at CMS~\cite{Sirunyan:2017qdw} experiment. }}
  \label{Fig:SigmaY3SCMS13TeV}
\end{figure}

\begin{table*}
  \centering
  \caption{Comparison of CS elements and CO LDMEs extracted from fitting with experimental data
    using NRQCD formalism for $\Upsilon$(3S).}
  \footnotesize
  \begin{tabular*}{\textwidth}{@{\extracolsep{\fill}}lrrrrrl@{}}
    \hline
    \hline
    Ref. (LO/NLO) &PDF & $m_b$ & $M_L(b\bar{b}([^3S_1]_1$ & $M_L(b\bar{b}([^3S_1]_8$ & 
    $M_L(b\bar{b}([^1S_0]_8$, & $p_T$-cut \\
    & & & $\rightarrow\Upsilon(3S)$ & $\rightarrow\Upsilon(3S)$ & $[^3P_0]_8\rightarrow\Upsilon(3S)$ & \\
    & & (GeV) & $({\rm GeV^3})$ & $({\rm GeV^3})$ & $({\rm GeV^3})$ & GeV/$c$ \\
    \hline
    \hline
    & & & & & & \\
    present (LO) & CT14LO &4.88 &4.3 & 0.0547$\pm$0.0007$\pm$0036 & 0.0054$\pm$0.0005$\pm$0.0021 & 8   \\
    & & & & & & \\
    \cite{Domenech:2000ri} (LO) & CTEQ4L & 4.88 & 3.54 & 0.099$\pm$0.011 & 0 & 2 \\
    & & & & 0.091$\pm$0.015 & 0 & 4 \\
    & & & & 0.068$\pm$0.011 & 0 & 8 \\
    & & & & & & \\
    \cite{Brateen:PRD2001} (LO) & CTEQ5L & 4.77 & 4.3$\pm$0.9 & 0.036$\pm$0.019 & 0.0108$\pm$0.0086 & 8 \\
    & & & & 0.039$\pm$0.017 & 0.0342$\pm$0.0276 & \\
    & & & & & & \\
    & MRSTLO & 4.77 & 4.3$\pm$0.9 & 0.037$\pm$0.021 & 0.0150$\pm$0.0098 & 8 \\
    & & & & 0.041$\pm$0.019 & 0.0474$\pm$0.0312 & \\
    & & & & & & \\
    \cite{Gong:2010bk} (NLO) & CTEQ6M & 5.18 & 1.128 & 0.03250$\pm$0.00876 & 0.000920$\pm$0.000968 & -\\
    & & & & & & \\
    \cite{Sharma:2012dy} (LO) & MSTW08LO & 4.88 & 4.3 & 0.0513$\pm$0.0085 & 0.0002$\pm$0.0062 & -  \\
    & & & & & & \\
    \cite{Gong:2013qka} (NLO) & CTEQ6M & 5.18 & 1.128 & 0.0271$\pm$0.0013 & 0.00956$\pm$0.00476 & 8 \\
    & & & & & & \\
    \cite{Feng:2015wka} (NLO) & CTEQ6M & 5.18 & 1.128 & 0.0132$\pm$0.0020 & -0.00520$\pm$0.00518 & 8 \\
    \hline
    \hline
  \end{tabular*}
  \label{Tab:LDMEsY3S}
\end{table*}
\normalsize

\begin{figure}
\centering
  \includegraphics[width=0.49\textwidth]{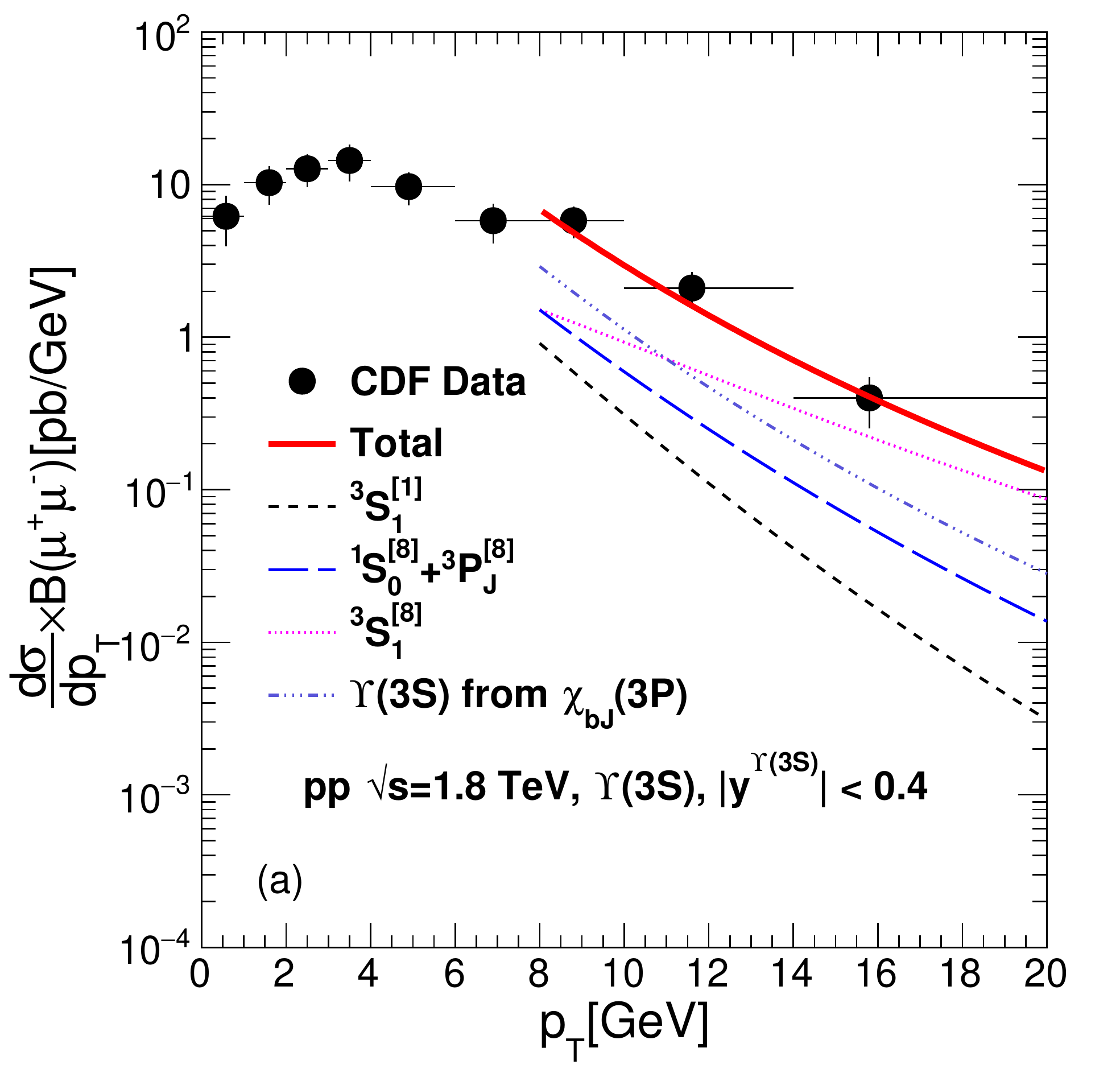}
  \includegraphics[width=0.49\textwidth]{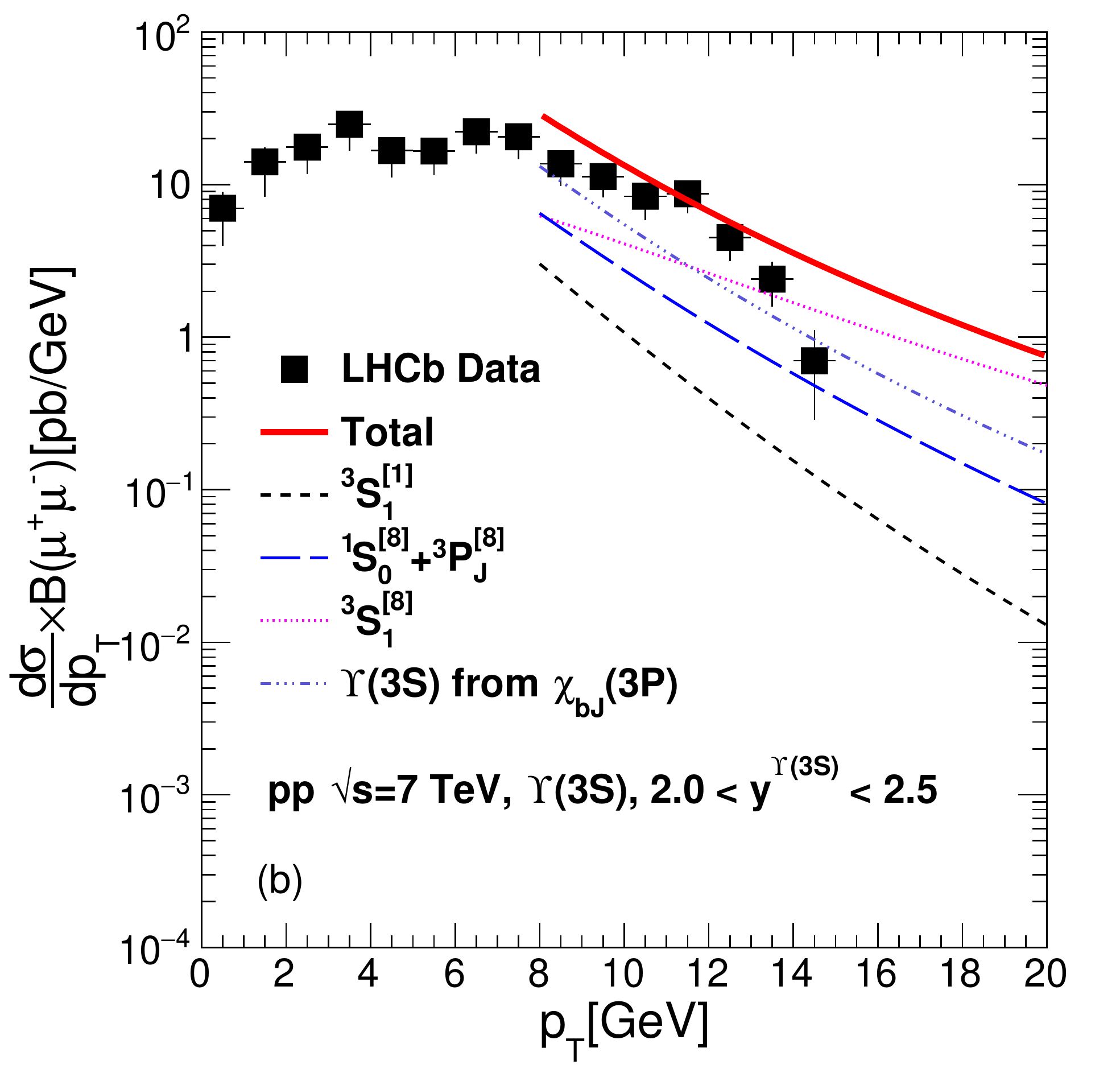} 
  \caption{\small{The NRQCD calculations of production cross-section of $\Upsilon$(3S) in
    p +{$\bar {\rm p}$} collisions at $\sqrt{s}$ = 1.8 TeV and p+p collisions at
    7 TeV in forward rapidities, as a function of transverse momentum compared with the measured data 
   at CDF~\cite{Acosta:2001gv} and LHCb~\cite{LHCb:2012aa} experiment. }}
  \label{Fig:SigmaY3SCDF}
\end{figure}


\begin{figure}
  \centering
  \includegraphics[width=0.49\textwidth]{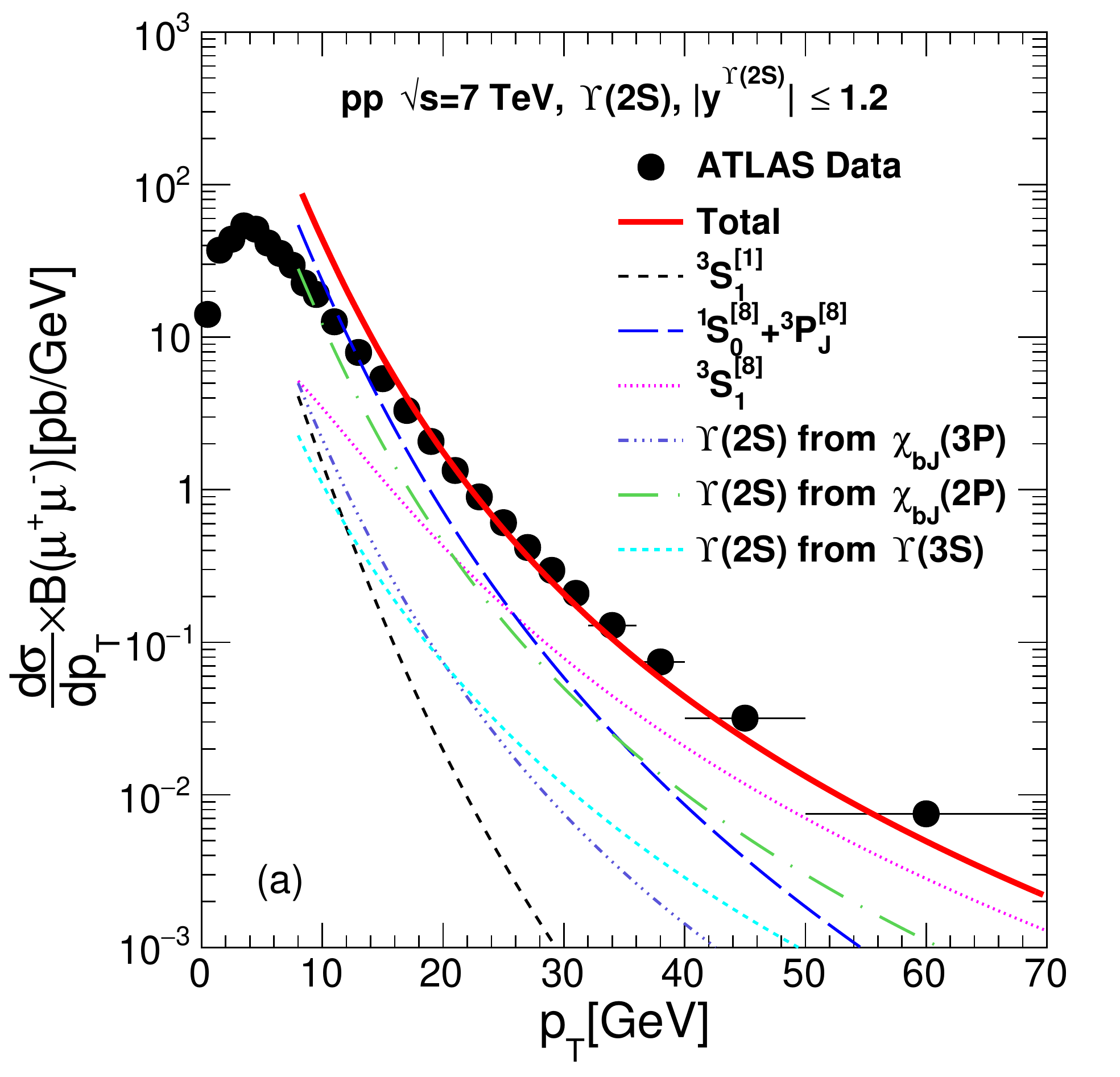}
  \includegraphics[width=0.49\textwidth]{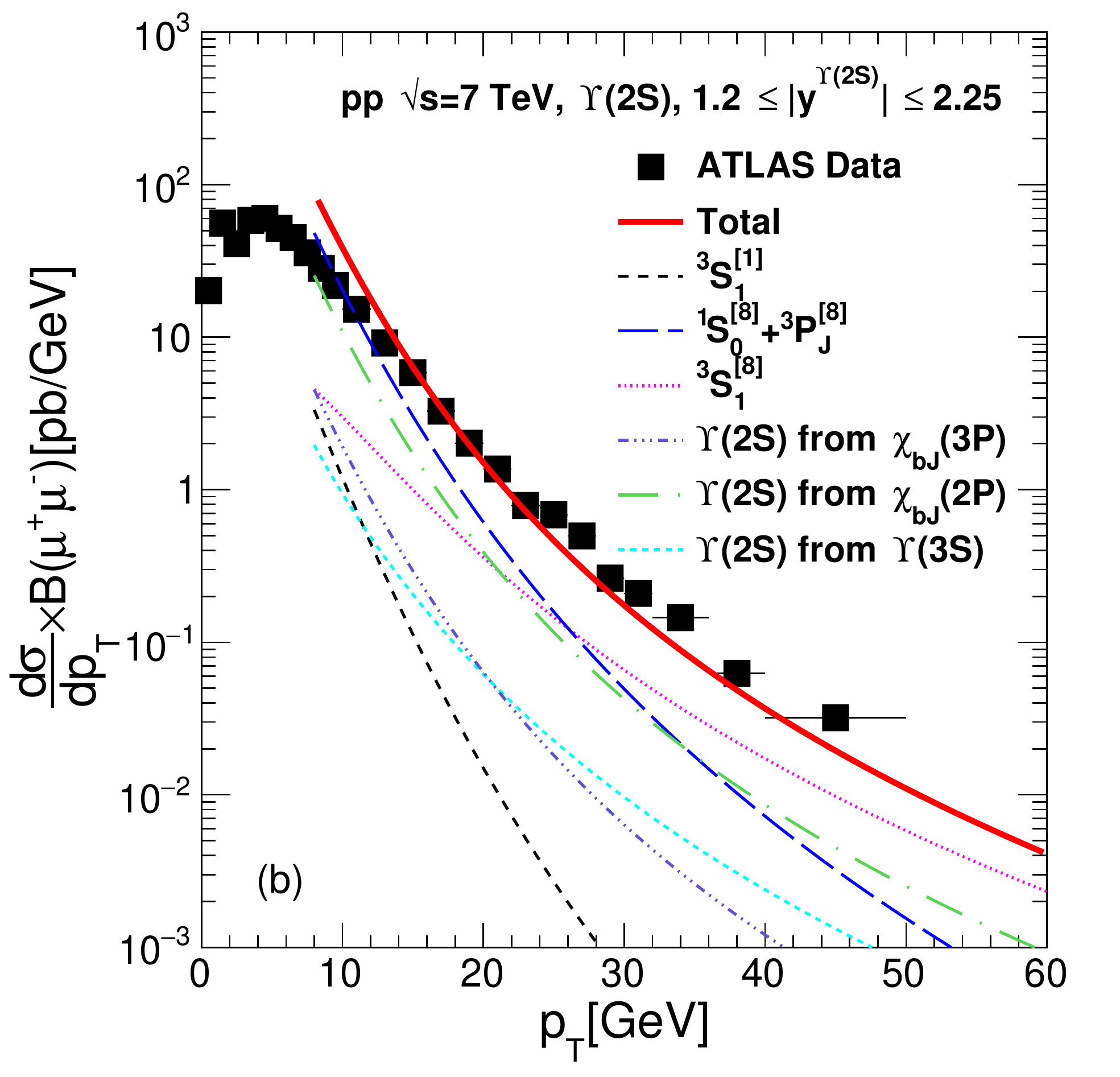}
  \includegraphics[width=0.49\textwidth]{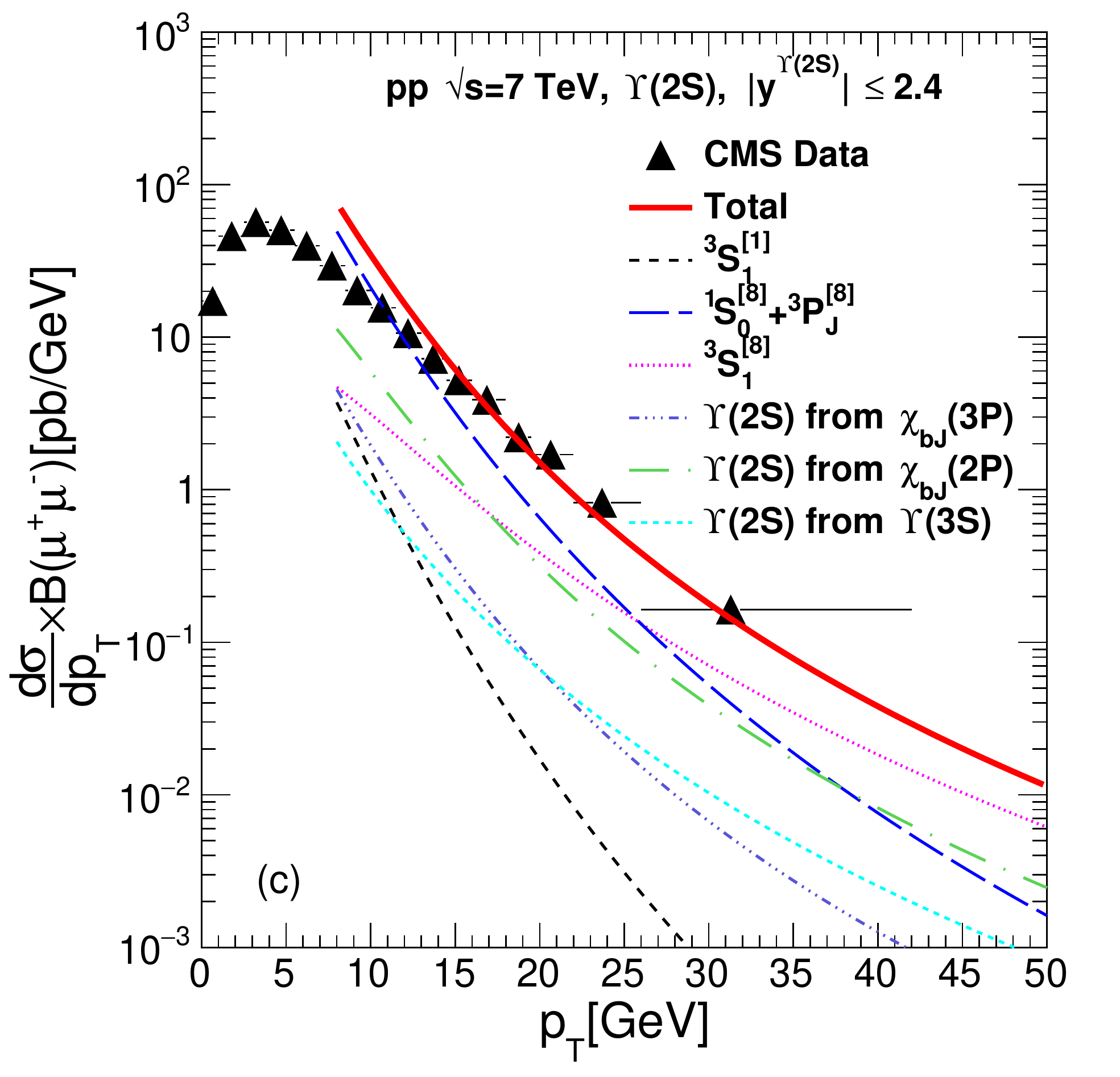}
  \caption{\small{The NRQCD calculations of production cross-section of $\Upsilon$(2S) in p+p collisions at 
    $\sqrt{s}$ = 7 TeV in central and forward rapidities, as a function of transverse momentum compared with the measured data 
    at CMS~\cite{Chatrchyan:2013yna} and ATLAS~\cite{Aad:2012dlq} experiments. }}
  \label{Fig:SigmaY2SATLAS}
\end{figure}

\begin{figure}
\centering
  \includegraphics[width=0.49\textwidth]{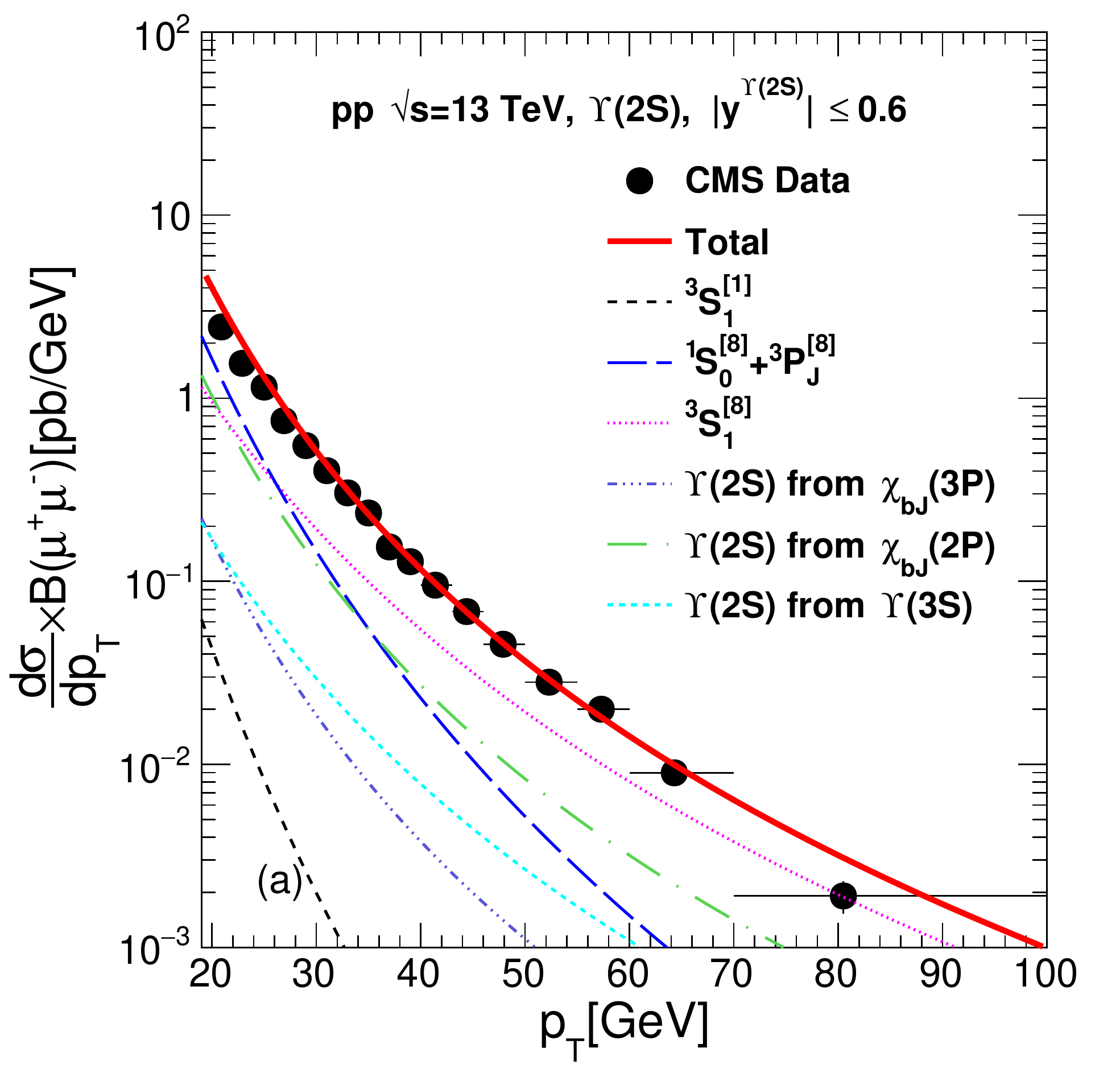}
  \includegraphics[width=0.49\textwidth]{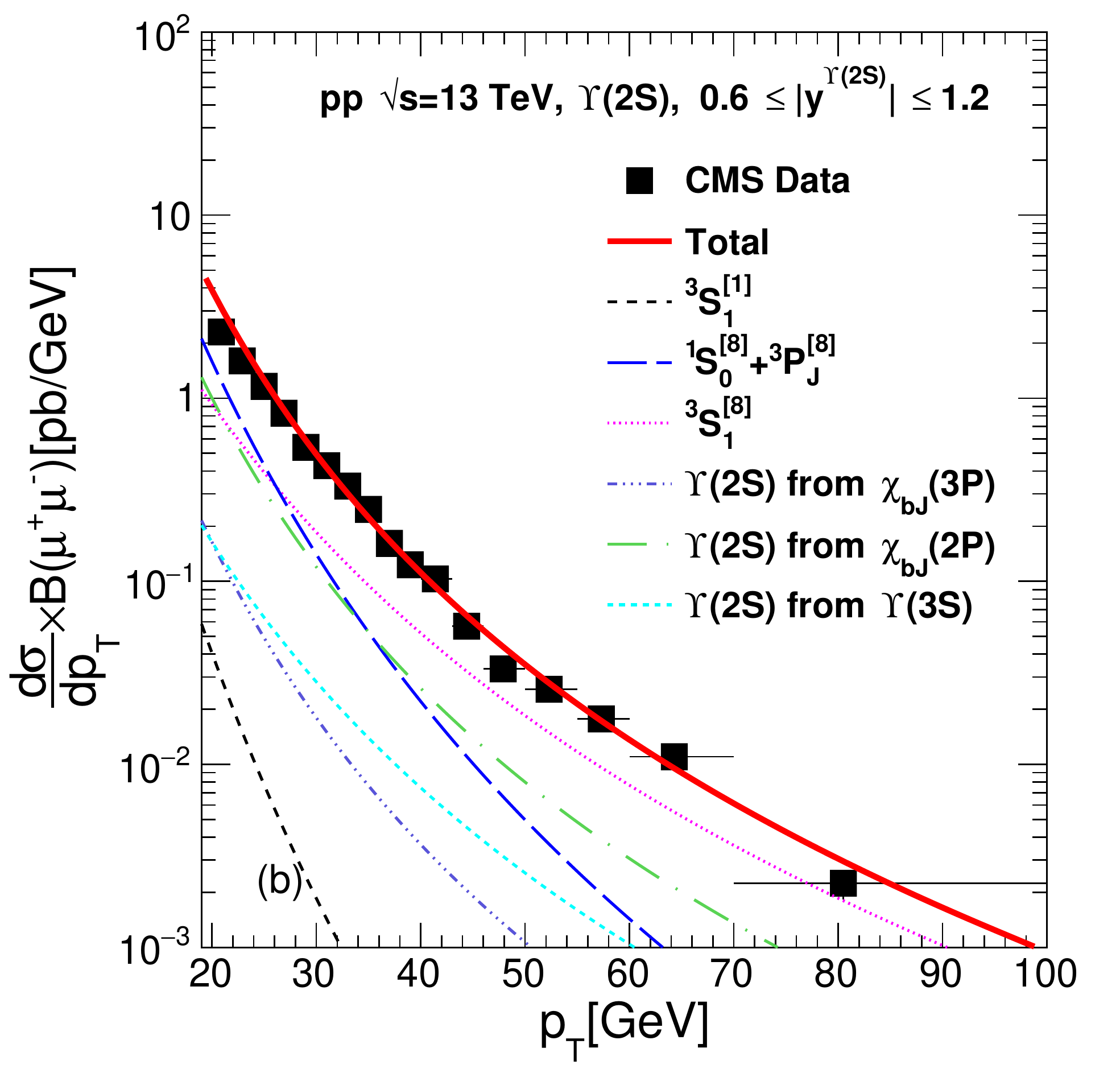} 
  \includegraphics[width=0.49\textwidth]{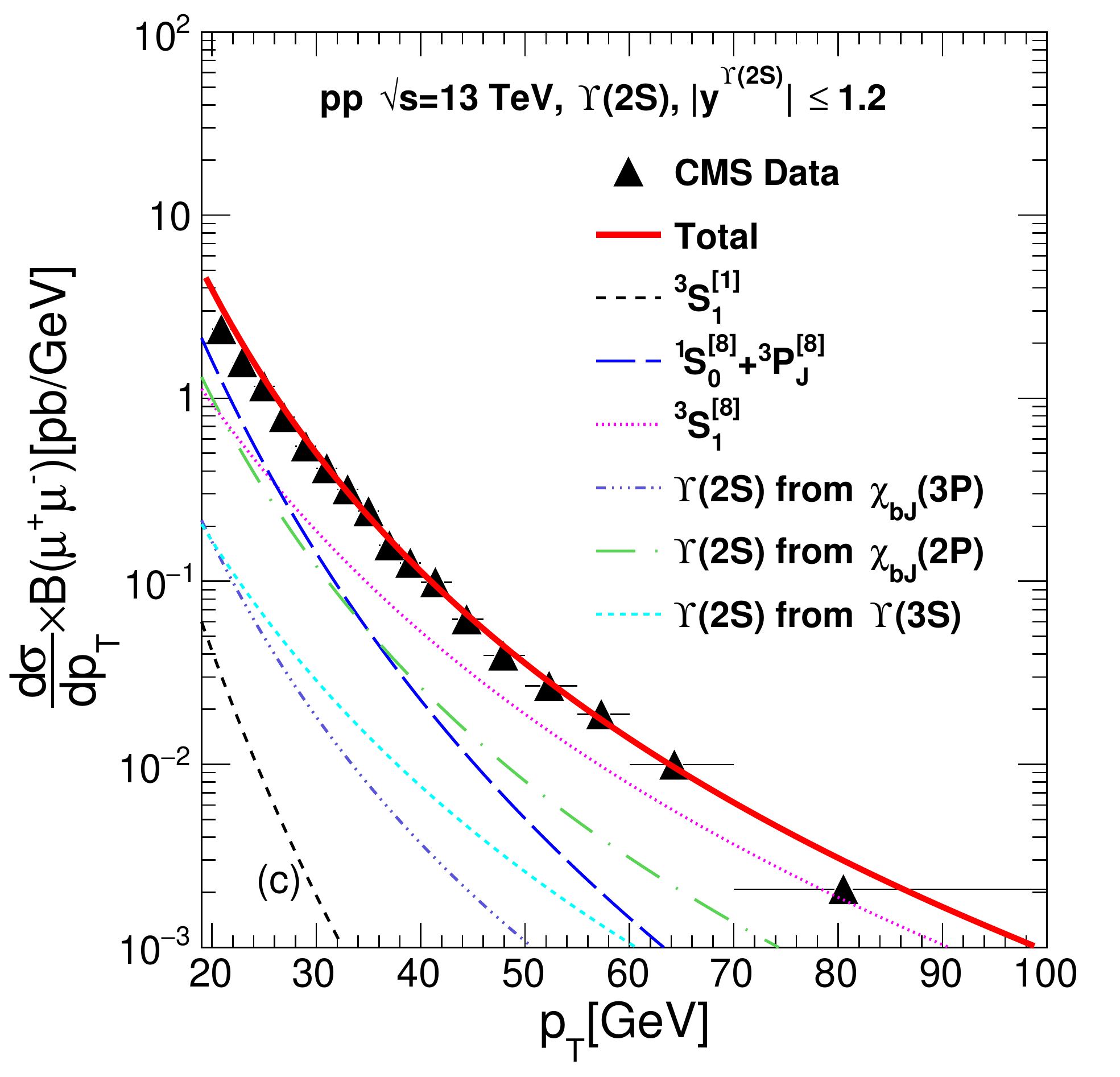}
 \caption{\small{The NRQCD calculations of production cross-section of $\Upsilon$(2S) in p+p collisions at 
   $\sqrt{s}$ = 13 TeV in central and forward rapidities, as a function of transverse momentum compared with the measured data 
   at CMS~\cite{Sirunyan:2017qdw} experiment. }}
  \label{Fig:SigmaY2SCMS13TeV}
\end{figure}

\begin{figure}
  \centering
  \includegraphics[width=0.49\textwidth]{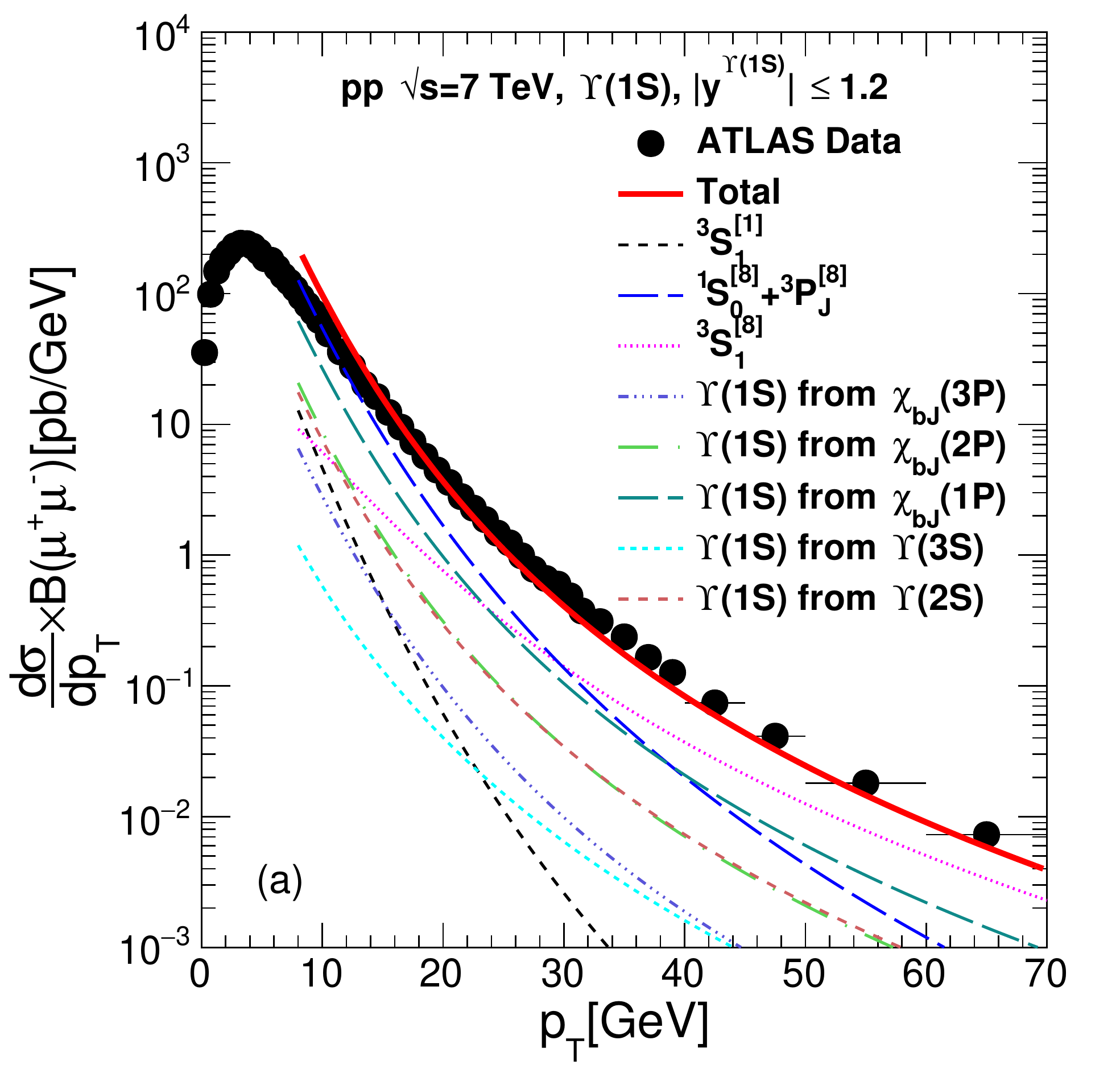}
  \includegraphics[width=0.49\textwidth]{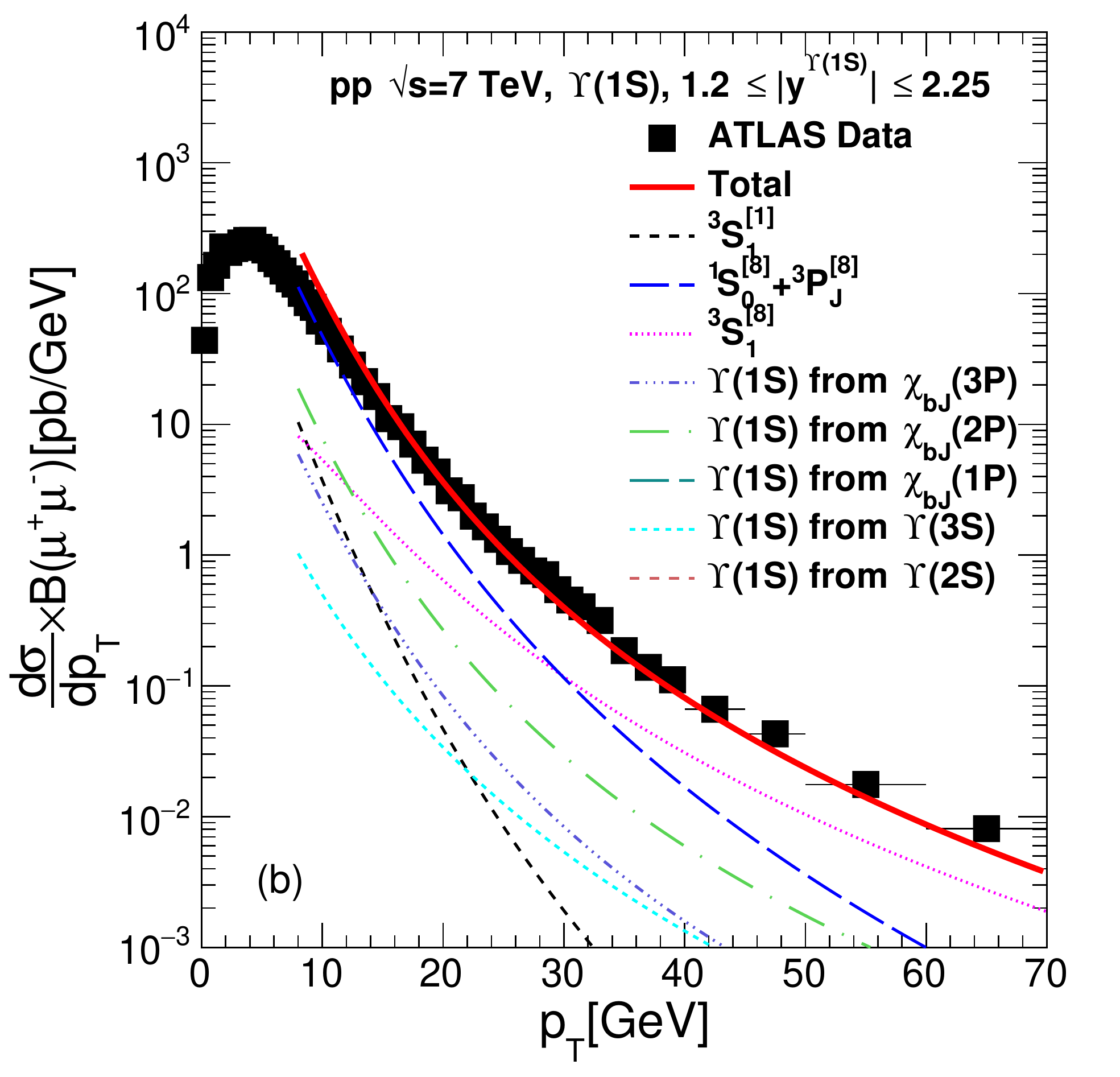}
  \includegraphics[width=0.49\textwidth]{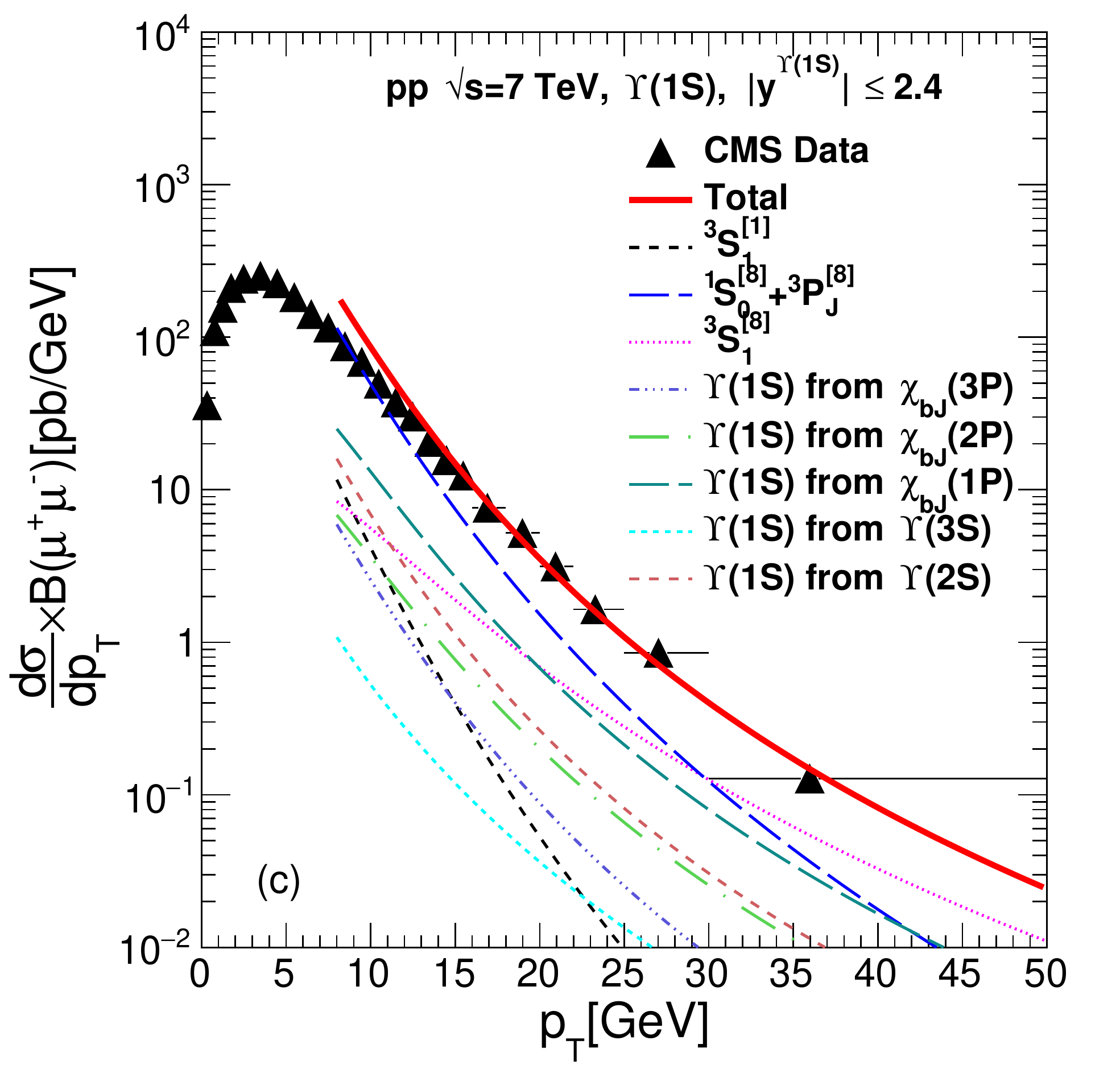}
 \caption{\small{The NRQCD calculations of production cross-section of $\Upsilon$(1S) in p+p collisions at 
   $\sqrt{s}$ = 7 TeV in central and forward rapidities, as a function of transverse momentum compared 
   with the measured data at ATLAS~\cite{Aad:2012dlq} and CMS~\cite{Chatrchyan:2013yna} experiments.}}
  \label{Fig:SigmaY1SATLAS7TEV}
\end{figure}

\begin{figure}
  \centering
  \includegraphics[width=0.49\textwidth]{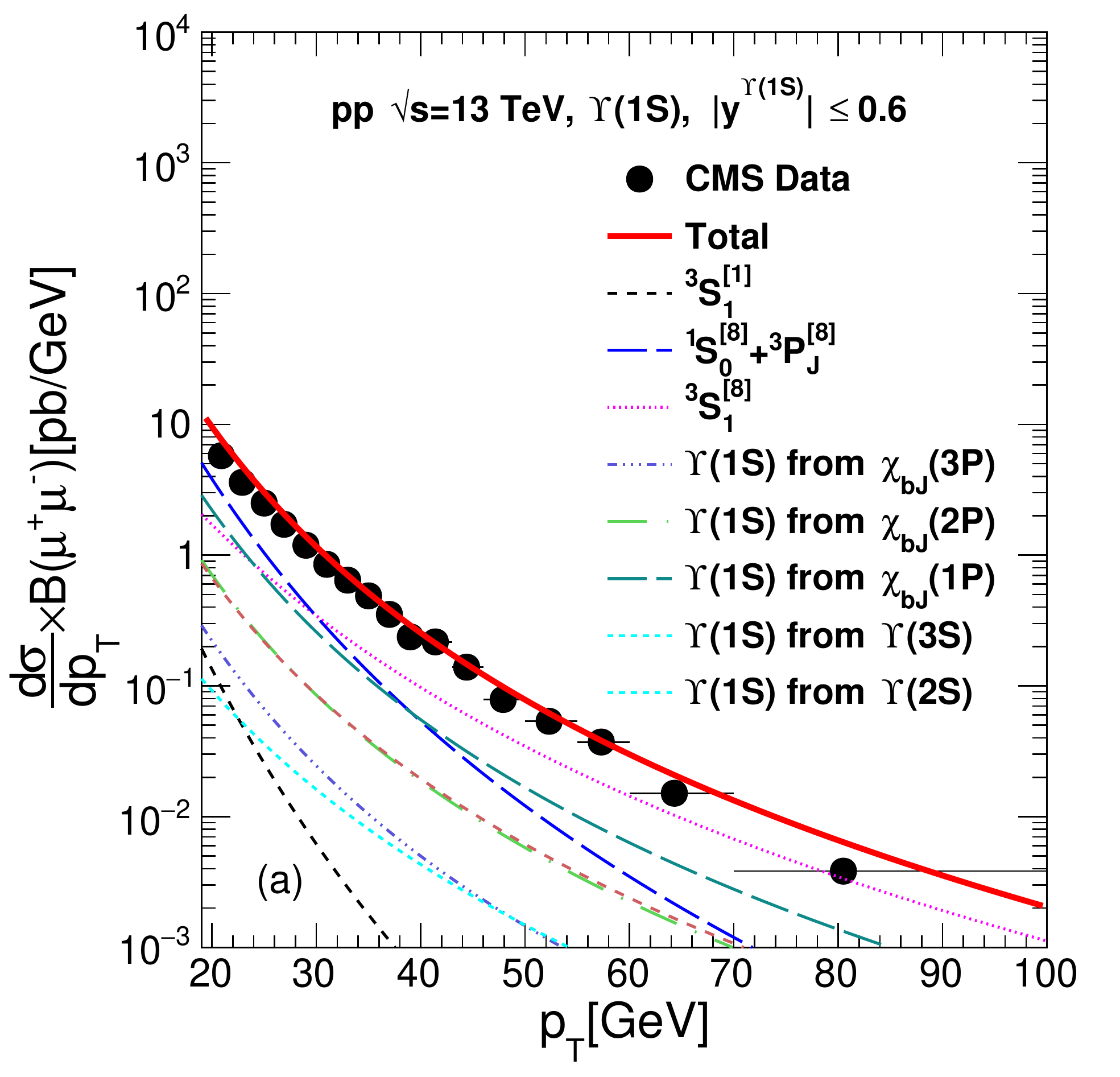}
  \includegraphics[width=0.49\textwidth]{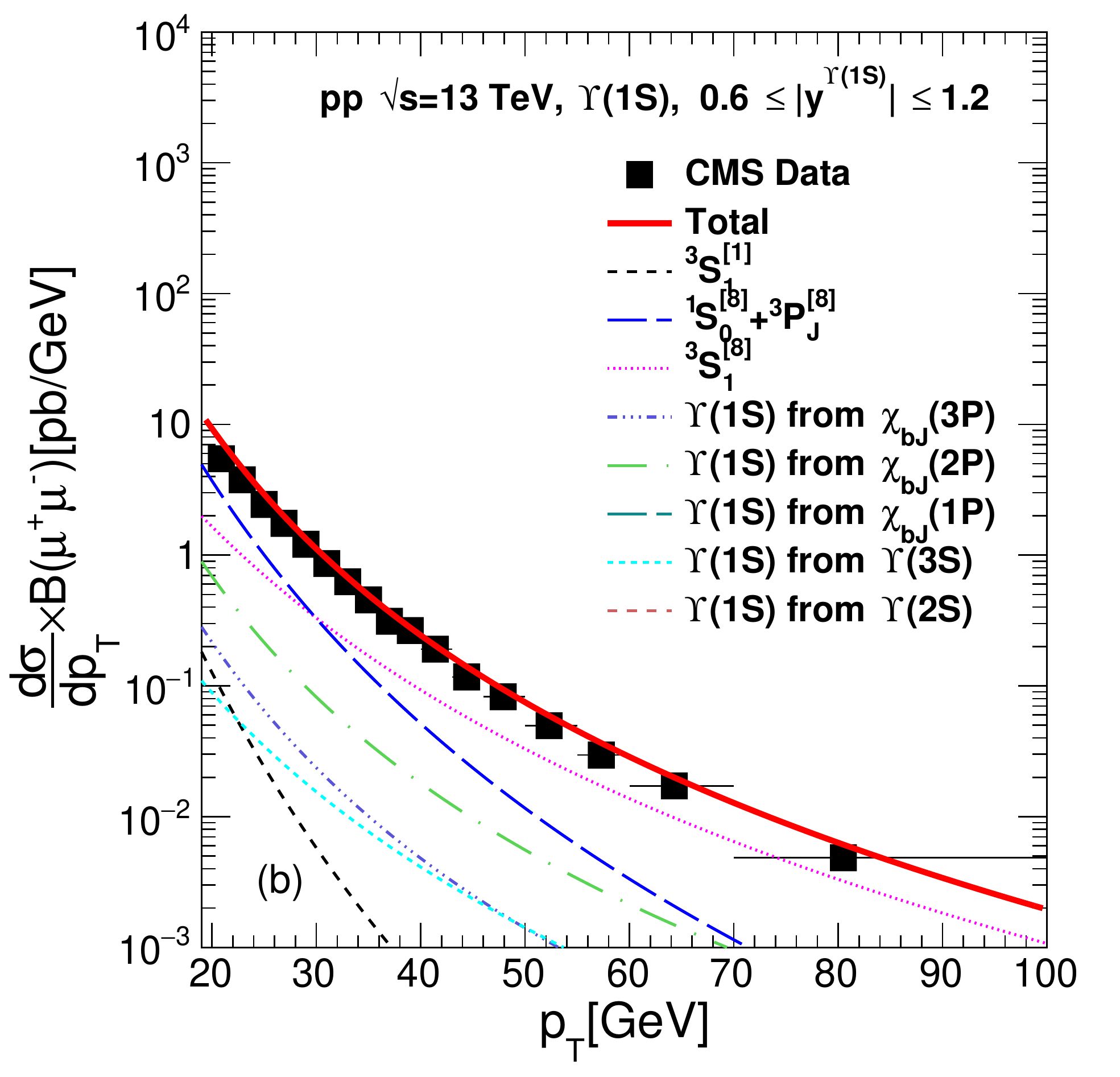} 
  \includegraphics[width=0.49\textwidth]{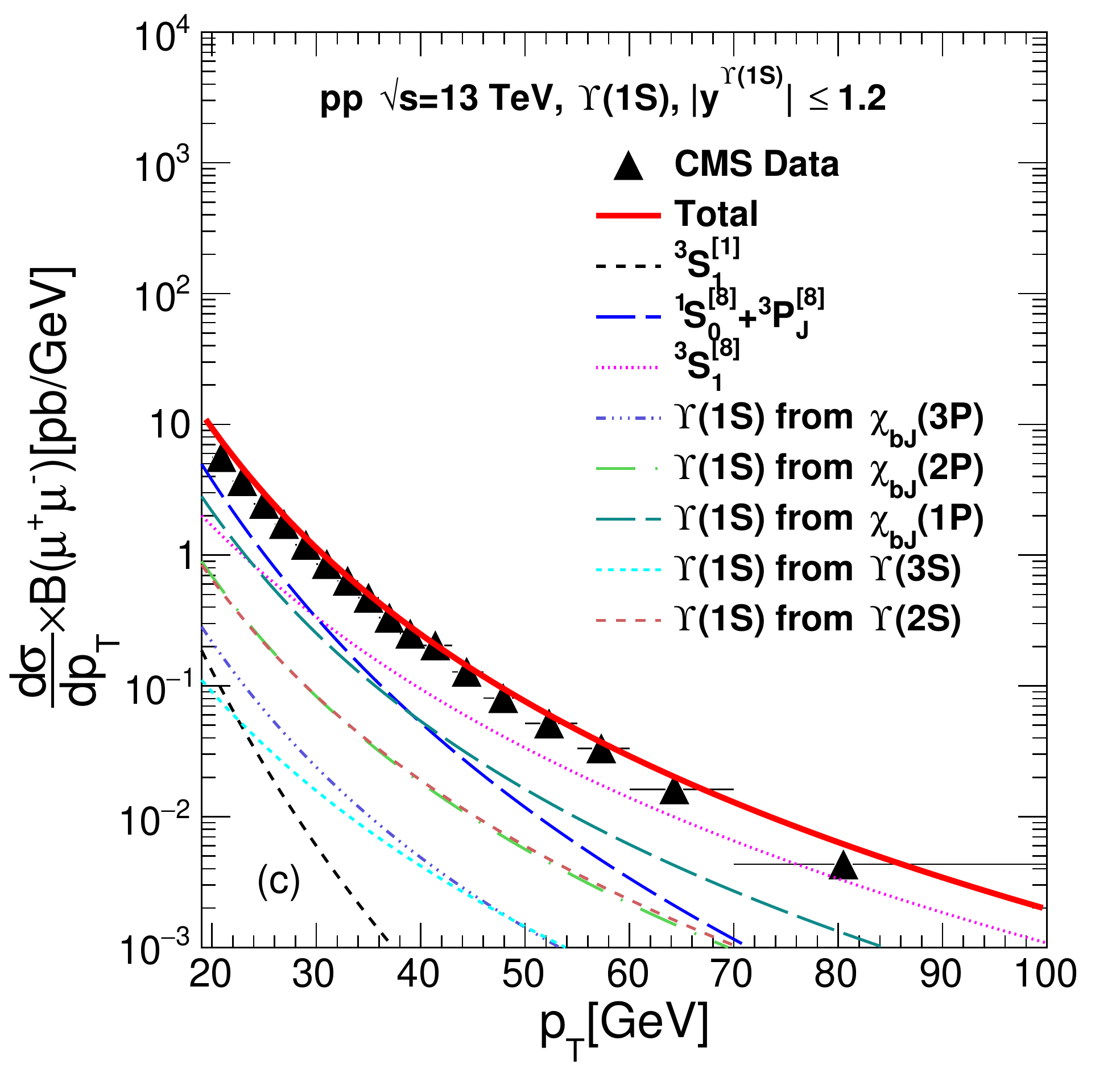}
 \caption{\small{The NRQCD calculations of production cross-section of $\Upsilon$(1S) in p+p collisions at 
   $\sqrt{s}$ = 13 TeV in central and forward rapidities, as a function of transverse momentum compared with the measured data 
   at CMS~\cite{Sirunyan:2017qdw} experiment. }}
  \label{Fig:SigmaY1SCMS13TeV}
\end{figure}

\section{Results and discussions}
\label{sec:results}
We first start with the production of $\Upsilon$(3S) which has feed down contributions
only from $\chi_{b}$(3P).
As described, the expressions and the values for the 
colour-singlet elements can be obtained by solving the non-relativistic 
wavefunctions~\cite{Cho:1995vh}. 
The CO LDMEs on the other hand, cannot be connected to the non-relativistic
wavefunctions of $b \bar b$. 
The measured data sets from different experimental collaborations are thus used to constrain them.

 Figure~\ref{Fig:SigmaY3SCMS} shows the NRQCD calculations of production cross-section of 
$\Upsilon$(3S) in p+p collisions as a function of transverse momentum compared with the 
measured data in CMS~\cite{Khachatryan:2015qpa} and ATLAS~\cite{Aad:2012dlq} 
detectors at LHC in central rapidities. {\color{black}Figure~\ref{Fig:SigmaY3SCMS} and all subsequent figures
are normalized by corresponding rapidity intervals.} In Figure~\ref{Fig:SigmaY3SCMS_forwardRap},
similar comparisons have been shown with data for $1.2<|y|<2.25$ and $|y|<2.4$
measured at ATLAS~\cite{Aad:2012dlq} and CMS~\cite{Chatrchyan:2013yna} detectors 
respectively. Figure~\ref{Fig:SigmaY3SCMS13TeV} corresponds to CMS ~\cite{Sirunyan:2017qdw}
measurements at $\sqrt{s}=13$ TeV for rapidities, $|y|<0.6$, $0.6<|y|<1.2$ and $|y|<1.2$, whereas in 
Figure~\ref{Fig:SigmaY3SCDF} we have used measurements from CDF~\cite{Acosta:2001gv}
collaboration in p +{$\bar {\rm p}$} at $\sqrt{s}=1.8$ TeV with $|y|<0.4$ as well as
that from LHCb ~\cite{LHCb:2012aa}
collaboration in p+p collisions at $\sqrt{s}=7$ TeV with rapidities $2.0<y<2.5$.
The LDMEs are obtained by a combined fit using all the aforesaid datasets.
The $\chi^2$/ndof is $\sim 4 $ for the combined fitting.
{\color{black} To estimate the uncertainty in the LDMEs following study is performed by varying two
  parameters of the calculation}
\begin{enumerate}[i]
  \color{black}
\item We extracted an estimate of uncertainties in the LDMEs due to enhancement of color singlet cross-section
  of quarkonia by around a factor of two expected from NLO corrections~\cite{Gong:2010bk,Gong:2008hk}.
\item We changed the mass of the bottom quark to 4.77 GeV from 4.88 GeV and recalculated the short distance cross-sections.
  LDMEs are then extracted using these cross-sections. This value of the mass is motivated by the use
  of several groups earlier in their calculations~\cite{Brateen:PRD2001,Gong:2013qka}. 
  \end{enumerate} 
    {\color{black} Both of these uncertainties are then added in quadrature and quoted with the LDME values.}
Table~\ref{Tab:LDMEsY3S} contains LDMEs for $\Upsilon$(3S) extracted in present analysis
in comparison with different other results. Here the first error is due to the fitting and
the second error is obtained by the uncertainty study. 
Our result for the matrix element $M_L(b\bar{b}([^3S_1]_8))$
shows a close proximity with LO analysis of Ref.~\cite{Brateen:PRD2001,Sharma:2012dy}.
 In our work, we have considered a linear combination of the other two colour octet 
LDMEs in the form of $\frac{M_{L}([^1S_0]_{8})}{5}+\frac{3M_{L}([^3P_0]_{8})}{m_b^2}$, same as that
done in Ref.~\cite{Sharma:2012dy}.
 There have been different ways to treat the colour octet LDMEs in the literature.
In Ref.~\cite{Domenech:2000ri}, 
the authors have taken this combination as $M_{L}([^1S_0]_{8})+\frac{5M_{L}([^3P_0]_{8})}{m_b^2}$.
In Ref.~\cite{Brateen:PRD2001}, these two matrix elements,
$M_{L}([^1S_0]_{8})$ and $\frac{5M_{L}([^3P_0]_{8})}{m_b^2}$
have been extracted separately using two different PDFs. In each case however, they have extracted
either of the two parameters considering the other to be vanishing.
The work in Ref.~\cite{Gong:2010bk} concentrates only on S-wave colour states.
In Refs.~\cite{Gong:2013qka,Feng:2015wka},
the parameters, $M_{L}([^1S_0]_{8})$ and $\frac{M_{L}([^3P_0]_{8})}{m_b^2}$ have been extracted
separately altogether. On the other hand in Ref.~\cite{Han:2014kxa}, the authors have
considered different combinations of 
colour octet states to fit with the experimental data with NRQCD at LO and NLO using 
CTEQ6L1 and CTEQ6M PDFs respectively with $m_b$=4.75 GeV and $[^3S_1]_1$=3.54 GeV$^3$.
Their extracted parameters are,
\begin{eqnarray}
  \nonumber
  M_{0,r_0}~=~[^1S_0]_8+\frac{r_0}{m_b^2}[^3P_0]_8~=~0.0283\pm 0.0007{\rm \,\, GeV^3} \nonumber \\
  M_{1,r_1}~=~[^3S_1]_8+\frac{r_1}{m_b^2}[^3P_0]_8~=~0.0083\pm 0.0002{\rm \,\,GeV^3} 
\end{eqnarray}
with $r_0$=3.8 and $r_1$=-0.52 GeV$^2$.

 After fixing the $\Upsilon$(3S) yield, we next consider $\Upsilon$(2S) production 
that has feed down contributions from $\Upsilon$(3S), $\chi_b(3P)$ and $\chi_b(2P)$ states
along with the direct production. The corresponding
branching fractions for the feed down sectors are given in
Table~\ref{BRUpsilon}. We have used our extracted values of the $\Upsilon$(3S) LDMEs
for the feed down contributions from the $\Upsilon$(3S). To include the $\chi_b(nP)$ states feed down
LDMEs are obtained from Ref.~\cite{Sharma:2012dy,Feng:2015wka}.

In Fig~\ref{Fig:SigmaY2SATLAS}, we have shown our NRQCD predictions of production
cross-sections for $\Upsilon$(2S) in p+p collisions as functions of 
$p_T$ along with the measured data in CMS~\cite{Chatrchyan:2013yna} and ATLAS~\cite{Aad:2012dlq}
detectors at central and forward rapidities. All the contributions
alongwith feed down ones are displayed separately. Fig.~\ref{Fig:SigmaY2SCMS13TeV} describes 
the same alongwith the data from CMS detector at 13 TeV for 
both central and forward rapidities. Our results of CO LDMEs for 
$\Upsilon$(2S) have been given in Table~\ref{LDMEsY2S} along with existing results
from different other groups.
Our value for $M_L(b\bar{b}([^3S_1]_8 \rightarrow \Upsilon(2S))$ is in agreement with the
values from other groups also $M_L(b\bar{b}([^1S_0]_8,[^3P_0]_8 \rightarrow \Upsilon(2S))$
does not have negative value (which is unphysical) unlike some other groups.
The inclusion of 13 TeV data along with the incorporation of feed down from $\chi_{b}$(3P),
is expected to give better constrains of LDMEs.

In~\cite{Domenech:2000ri,Brateen:PRD2001,Sharma:2012dy,Gong:2013qka,Feng:2015wka,Han:2014kxa}, 
authors have considered different combinations of
CO LDMEs that has already been described. In Ref.~\cite{Han:2014kxa}, the extracted 
parameters for $\Upsilon$(2S) are,
\begin{eqnarray}
\nonumber
M_{0,r_0}=0.0607\pm0.0108 \,\,{\rm GeV^3}\\ \nonumber
M_{1,r_1}=0.0108\pm0.0020 \,\,{\rm GeV^3}
\end{eqnarray}
with $[3S_1]_1$=4.63 GeV$^3$ and the values of $r_0$ and $r_1$ are same as given before.
The $\chi^2$/ndof for the combined fit in our analysis is $\sim$ 3.

\begin{table*}
  \centering
  \caption{Comparison of CS elements and CO LDMEs extracted from fitting with experimental data
    using NRQCD formalism for $\Upsilon$(2S).}
  \footnotesize
  \begin{tabular*}{\textwidth}{@{\extracolsep{\fill}}lrrrrrl@{}}
    \hline
    \hline
    Ref. (LO/NLO) & PDF & $m_b$ & $M_L(b\bar{b}([^3S_1]_1$ & $M_L(b\bar{b}([^3S_1]_8$ & 
    $M_L(b\bar{b}([^1S_0]_8$, & $p_T$-cut \\
    & & & $\rightarrow\Upsilon(2S)$ & $\rightarrow\Upsilon(2S)$ & $[^3P_0]_8\rightarrow\Upsilon(2S)$ & \\
    & & (GeV) & $({\rm GeV^3})$ & $({\rm GeV^3})$ & $({\rm GeV^3})$ & GeV/$c$ \\
    \hline
    \hline
    & & & & & & \\
    present (LO) & CT14LO &4.88 &4.5 &0.0400$\pm$0.0016$\pm$0.0023 & 0.0405$\pm$0.0018$\pm$0.0029 & 8   \\
    & & & & & & \\
    \cite{Domenech:2000ri} (LO) & CTEQ4L & 4.88 & 5.01 & 0.040$\pm$0.029 & 0 & 2 \\
    & & & & 0.073$\pm$0.018 & 0 & 4 \\
    & & & & 0.103$\pm$0.027 & 0 & 8 \\
    & & & & & & \\
    \cite{Brateen:PRD2001} (LO) & CTEQ5L & 4.77 & 5.0$\pm$0.7 & 0.180$\pm$0.056 & -0.102$\pm$0.097 & 8 \\
    & & & & 0.172$\pm$0.050 & -0.106$\pm$0.102 & \\
    & & & & & & \\
    & MRSTLO & 4.77 & 5.0$\pm$0.7 & 0.196$\pm$0.063 & -0.087$\pm$0.111 & 8 \\
    & & & & 0.190$\pm$0.056 & -0.089$\pm$0.117 & \\
    & & & & & & \\
    \cite{Sharma:2012dy} (LO) & MSTW08LO & 4.88 & 4.5 & 0.0224$\pm$0.0200 & -0.0067$\pm$0.0084 & -  \\
    & & & & & & \\
    \cite{Gong:2013qka} (NLO) & CTEQ6M & 5.01 & 4.63 & 0.0030$\pm$0.0078 & 0.0075$\pm$0.0217 & 8 \\
    & & & & & & \\
    \cite{Feng:2015wka} (NLO) & CTEQ6M & 5.01 & 4.63 & 0.0222$\pm$0.0024 & -0.0003$\pm$0.0203 & 8 \\
    \hline
    \hline
  \end{tabular*}
  \label{LDMEsY2S}
\end{table*}
\normalsize
Having completed $\Upsilon$(3S) and $\Upsilon$(2S) parts, we now move on to explore $\Upsilon$(1S). 
Alongwith the direct yield, it has feed down contributions from higher
S-wave states like $\Upsilon$(3S) and $\Upsilon$(2S), as well as P-wave states like $\chi_b$(3P), $\chi_b$(2P)
and $\chi_b$(1P). The associated branching functions are provided in Table~\ref{BRUpsilon}.
The extracted CO-LDMEs for $\Upsilon$(3S) and $\Upsilon$(2S) are used for feed down contributions,
whereas the LDMEs for the $\chi_b$(nP) states have been taken from Ref.~\cite{Sharma:2012dy,Feng:2015wka}
for this present case study.
In Fig.~\ref{Fig:SigmaY1SATLAS7TEV}, we have displayed our NRQCD calculation of production cross-section 
of $\Upsilon$(1S) as function of $p_T$ along with the experimental measurements by ATLAS and
CMS at $\sqrt{s}$=7 TeV in central rapidities.
Finally in Fig.~\ref{Fig:SigmaY1SCMS13TeV}, we present our results along with
the CMS measurements at 13 TeV with all the components separately to signify their relative
contributions. 
\begin{table*}
  \centering
  \caption{Comparison of CS elements and CO LDMEs extracted from fitting with experimental data
    using NRQCD formalism for $\Upsilon$(1S).}
  \footnotesize
  \begin{tabular*}{\textwidth}{@{\extracolsep{\fill}}lrrrrrl@{}}
    \hline
    \hline
    Ref. (LO/NLO) & PDF & $m_b$ & $M_L(b\bar{b}([^3S_1]_1$ & $M_L(b\bar{b}([^3S_1]_8$ & 
    $M_L(b\bar{b}([^1S_0]_8$, & $p_T$-cut \\
    & & & $\rightarrow\Upsilon(1S)$ & $\rightarrow\Upsilon(1S)$ & $[^3P_0]_8\rightarrow\Upsilon(1S)$ & \\
    & & (GeV) & $({\rm GeV^3})$ & $({\rm GeV^3})$ & $({\rm GeV^3})$ & GeV/$c$ \\
    \hline
    \hline
    & & & & & & \\
    present (LO) & CT14LO &4.88 &10.9 &0.0556$\pm$0.0017$\pm$0.0030 & 0.0735$\pm$0.0016$\pm$0.0060 & 8   \\
    & & & & & & \\
    \cite{Domenech:2000ri} (LO) & CTEQ4L & 4.88 & 11.1 & 0.077$\pm$0.017 & 0 & 2 \\
    & & & & 0.087$\pm$0.016 & 0 & 4 \\
    & & & & 0.106$\pm$0.013 & 0 & 8 \\
    & & & & & & \\
    \cite{Brateen:PRD2001} (LO) & CTEQ5L & 4.77 & 12.8$\pm$1.6 & 0.116$\pm$0.027 & 0.109$\pm$0.062 & 8 \\
    & & & & 0.124$\pm$0.025 & 0.111$\pm$0.065 & \\
    & & & & & & \\
    & MRSTLO & 4.77 & 12.8$\pm$1.6 & 0.117$\pm$0.030 & 0.181$\pm$0.072 & 8 \\
    & & & & 0.130$\pm$0.028 & 0.186$\pm$0.075 & \\
    & & & & & & \\
    \cite{Sharma:2012dy} (LO) & MSTW08LO & 4.88 & 10.9 & 0.0477$\pm$0.0334 & 0.0121$\pm$0.0400 & -  \\
    & & & & & & \\
    \cite{Gong:2013qka} (NLO) & CTEQ6M & 4.75 & 9.282 & -0.0041$\pm$0.0024 & 0.0780$\pm$0.0043 & 8 \\
    & & & & & & \\
    \cite{Feng:2015wka} (NLO) & CTEQ6M & PDG & 9.282 & 0.0061$\pm$0.0024 & 0.0895$\pm$0.0248 & 8 \\
    \hline
    \hline
  \end{tabular*}
  \label{LDMEsY1S}
\end{table*}
\normalsize

Table~\ref{LDMEsY1S} shows our results for $\Upsilon$(1S) parameters along with
the results from different groups. The individual values of LDMEs are in agreement with
the values from previous works but with considerable reduction in 
errors upon inclusion of 13 TeV data sets from CMS.
The values of the parameters $M_{0,r_0}$ and $M_{1,r_1}$ extracted in Ref.~\cite{Han:2014kxa} are,
\begin{eqnarray}
\nonumber
M_{0,r_0}=0.1370\pm0.0111 \, {\rm GeV^3},\\ \nonumber
M_{1,r_1}=0.0117\pm0.0002 \, {\rm GeV^3}
\end{eqnarray}
with $[3S_1]_1$=9.28 GeV$^3$ keeping $r_0$ and $r_1$ same as given before.

\section{Summary}
\label{sec:summary}
We have presented NRQCD calculations for the differential production 
cross-sections of $\Upsilon$ states in  p+p collisions.  Measured transverse momentum
distributions of $\Upsilon$(3S), 
$\Upsilon$(2S) and $\Upsilon$(1S) in p +{$\bar {\rm p}$} collisions at $\sqrt{s}=$ 1.8 TeV and in 
p+p collisions at 7 TeV and 13 TeV are used to constrain the LDMEs. All the relevant feeddown
contributions from higher mass states including the $\chi_{b}$(3P) are taken in to account.
The calculations for  $\Upsilon$(3S), $\Upsilon$(2S) and $\Upsilon$(1S) are compared with 
the measured data at Tevatron and LHC. The formalism provides  very good description of the data in 
large transverse momentum range at different collision energy. 
We compare the LDMEs for bottomonia obtained in this analysis with the results from earlier works.
At high $p_T$, the colour singlet contribution is very small and LHC data in large $p_T$ range 
help to constrain the relative contributions of different colour octet contributions.
For $\Upsilon$ states at high  $p_T$, the contribution of the  
$M_{L}(b\barb([^3S_1]_{8}\rightarrow \Upsilon(nS)))$ is highest which is opposite to the charmonia 
case where the contribution for the combination of  $M_{L}(c\barc([^1S_0]_{8},[^3P_0]_{8})\rightarrow \psi)$ 
is more~\cite{Kumar:2016ojy}.
In summary, we present a comprehensive lowest-order analysis of hadroproduction data of bottomonia 
states using the latest parton distribution functions and including very recent LHC data. The feed-down
contributions from all the $\chi_{b}$ states are included in the calculations. The values of relevant LDMEs
are extracted by doing a simultaneous fit of all the data sets. These values will be useful for predictions
of quarkonia cross-section and for the purpose of a comparison with those obtained using the NLO formulations.

\section*{acknowledgement}
Authors thank Board of Research in Nuclear Sciences (BRNS) and UGC (DRS) for support.
AB thanks Alexander von Humboldt (AvH) foundation and Federal Ministry
of Education and Research (Germany) for support through Research Group
Linkage Programme. KS acknowledges the financial support from DST-SERB under 
NPDF file no. PDF/2017/002399.

\end{document}